\documentclass[%
 reprint,
 amsmath,amssymb,
 aps,
]{revtex4-2}

\usepackage{graphicx}
\usepackage{dcolumn}
\usepackage{bm}

\begin{document}

\preprint{APS/123-QED}

\title{Single pixel wide gamut dynamic color modulation based on \\graphene micromechanical system}

\author{Yanli Xu$^{1,3}$}
\author{Hongxu Li$^2$}%
\author{Xin Zhang$^1$}
\author{Wenjing Liu$^3$}
\author{Zhengping Zhang$^1$}
\author{Shuijie Qin$^2$}
\email{shuijieqin@163.com}
\author{Jiangtao Liu$^3$}
\email{jtliu@semi.ac.cn}



\affiliation{$^1$College of Big Data and Information Engineering, Guizhou University, Guiyang 550025, P.R.China}
\affiliation{$^2$Key Laboratory of Photoelectric Technology and Application of Guizhou Province, Guizhou University, Guiyang 550025, P.R.China}%
\affiliation{$^3$College of Mechanical and Electrical Engineering, Guizhou Minzu University, Guiyang 550025, P.R.China}%
\affiliation{$^4$Guizhou Light Industry Technical College, Guiyang 550025, P.R.China}%



\date{\today}

\begin{abstract}
Dynamic color modulation in the composite structure of graphene microelectromechanical systems (MEMS)-photonic crystal microcavity is investigated in this work. The designed photonic crystal microcavity has three resonant standing wave modes corresponding to the three primary colors of red (R), green (G) and blue (B), forming strong localization of light in three modes at different positions of the microcavity. Once graphene is added, it can govern the transmittance of three modes. When graphene is located in the abdomen of the standing wave, which has strong light absorption and therefore the structure's transmittance is lower, or when graphene is located in the node of the standing wave, it has weak light absorption and therefore the structure's transmittance is higher. Therefore, the graphene absorption of different colors of light can be regulated dynamically by applying voltages to tune the equilibrium position of the graphene MEMS in the microcavity, consequently realizing the output of vivid monochromatic light or multiple mixed colors of light within a single pixel, thus greatly improving the resolution. Our work provides a route to dynamic color modulation with graphene and provides guidance for the design and manufacture of ultrahigh resolution, ultrafast modulation and wide color gamut interferometric modulator displays. 
\end{abstract}

\maketitle


\section{Introduction}

Among many nanomaterials, two-dimensional (2D) materials have emerged as promising candidates for mechanical and optoelectronic applications \cite{Wu2020,Li_2018,PhysRevB.102.085410}. Especially, graphene with an atomically thin structure (atom\-layers distance of $\sim$0.335 nm) exhibits a higher rigidity than steel (specific surface area of 2500 m$^{2}$/g), a stronger conductivity than copper  \cite{Fang2007Ab,2008Ultrahigh,DU2020108315}, remarkable mechanical and photoelectric properties (young's modulus of up to $\sim$1 TPa and charge-carrier mobility up to 20000 cm$^{2}$v$^{-1}$s$^{-1}$)  \cite{C6SM00108D,2017Mechanical,2008Measurement,D0NR07606F,PhysRevApplied.14.014056}, and a special experiment ductility (20\%) \cite{2015Observation,GUO2019108029}, thus making it an ideal candidate for micro- and nanoelectromechanical systems (MEMS and NEMS)  \cite{2015Graphene,Zhenyun2015Graphene,XugeFan2020Manufacture,PhysRevA.88.052521}. In addition, the suspended graphene material eliminates the interaction with the substrate, and give them the freedom of movement, which makes it has been widely used in applications of MEMS devices \cite{2015Mechanics,2008Imaging,CHASTE2017162,Legrand:17,PhysRevApplied.2.054008}, such as mechanical resonators \cite{2014Graphene,2014Optomechanical}, high performance sensors \cite{2013Electromechanical,2016Graphene,2016Low,C9NA00788A,doi:10.1021/acsami.7b17487,C5NR08668J}, electronic switches \cite{Nagase_2013,Sun2016Locally}, microphones \cite{Zhou2015Graphene}, and high resolution displays \cite{Santiago2018Graphene,doi:10.1021/acs.nanolett.6b02416,2017Very}.

In 2016, Dejan Davidovikj $et$ $al.$ visualized the motion of micrometer\-scale graphene nanodrums at ultrahigh frequencies with spatial resolution using a phase-sensitive interferometer \cite{2016Visualizing}. Meanwhile, Samer Houri $et$ $al.$ found that the double-layer graphene (DLG) membranes recreated the interference effect of interferometer modulation displays \cite{doi:10.1021/acs.nanolett.6b02416}. In 2018, they further reported a Graphene Interferometric Modulator Display (GIMD) prototype with 5$\mu$m-in-diameter pixels composed a high-resolution image (2500 pixels per inch), whose color can be changed at frequencies up to 400 Hz \cite{Santiago2018Graphene}. The results showed that the color reproducibility and speed of graphene membranes under a certain electro-optic modulation makes them suitable as pixels for high refresh rate displays. In recent years, display devices are developing diversely. With the increasing popularity of Virtual Reality, ultra-thin, fast-speed and high-resolution display technology is becoming the mainstream of current development \cite{Yoo:20,2017Continuous,D0NR07347D}. Obviously, the displays based on graphene mechanical pixels have above advantages and are more durable, energy efficient, flexible and easy to control than LED screens  \cite{Santiago2018Graphene,doi:10.1021/acs.nanolett.6b02416,2017Very,2012In}.

In this work, we studied the dynamic color modulation in the composite structure of graphene MEMS-photonic crystal microcavity. First of all, we designed a symmetric photonic crystal microcavity with three resonant standing wave modes corresponding to the three primary colors of red (R), green (G) and blue (B). The localized light fields are formed in three modes at different positions of the microcavity. Then, when graphene MEMS is coupled with photonic crystals, we found that graphene has a strong absorption when it is located in the  abdomen of the standing wave, and therefore the transmittance of the structure is lower, whereas graphene has the opposite property when it is located in the node of the standing wave. Therefore, the graphene absorption of different colors of light can be regulated dynamically by applying voltages to tune the equilibrium position of the graphene MEMS in the microcavity, thus realizing the output of vivid monochromatic light or multiple mixed colors of light within a single pixel. Traditionally, a single pixel of a Liquid Crystal Display (LCD) screen is actually composed of RGB sub-pixels, the superposition of which comprise what is known as a single pixel. It means that an LCD screen with a horizontal resolution of N whole pixels is actually composed of a linear array of 3N monochromatic sub-pixels, setting obstacles to manufacture higher-resolution displays. Thus, multiple mixed colors of light within a single pixel can greatly improve the resolution and reduce the power consumption of the display.

\section{Theoretical model and Method}

\begin{figure}
	\centering
	\includegraphics[width=\linewidth]{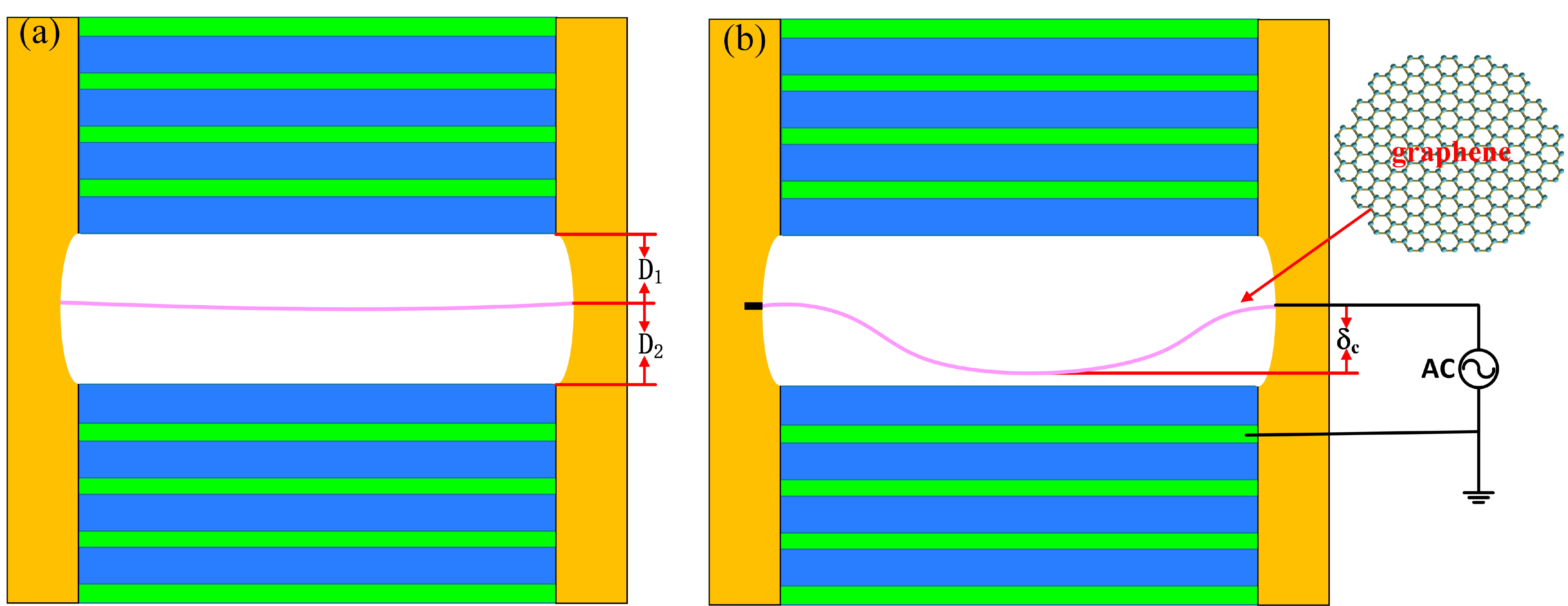}
	\caption{Structure of suspended graphene/optical microcavity ($D_1$ = 70 nm and $D_2$ = 1010 nm); (a) No voltage;(b) The membrane can be actuated electrostatically ($V_{ext}$$>$0 V). The electrodes are connected to graphene and the conductive layer in the Bragg mirror below. (The orange columns are insulated supports, and the inset shows the structure of graphene).}
	\label{Fig1}
\end{figure}

The schematic diagram of photonic crystal microcavity is shown in Fig. \ref{Fig1}. The cavity is formed by two Bragg mirrors and an air gap. The distributed Bragg mirrors consists of the ZnS layer and the ${\rm MgF_2}$ layer represented by green and blue, respectively. These two materials are alternately distributed on both sides of the microcavity with a period M = 4. The refractive indices of the two medium layer in the calculation are taken as n$_1$ = 2.5 and n$_2$ = 1.38, respectively. All layers are non-magnetic ($\mu=1$). The thicknesses of the corresponding layers are d$_1$ = $\lambda_0/4n_1$ and d$_2$ = $\lambda_0/4n_2$, where $\lambda_0$ = 530 nm. The fabrication of similar air gap photonic crystals has been quite mature \cite{2002Potential,1182782,1250494,Irmer_2005}, which provides a reasonable guarantee for the experimental implementation of our scheme. The light is bounced back and forth between the Bragg mirrors repeatedly in this microcavity, forming a standing wave distribution and increasing the transmission. Multiple resonance modes appear in the visible band once the cavity is long.

The deflection of graphene pixels can be used as a function of voltage, and the deflection can be used to correlate the electromechanical and ophotomechanical properties of the pixels \cite{Santiago2018Graphene}. We describe the pixels’ deflection by an axisymmetric parabolic profile given by $\delta(r)=\delta_c[1-(r/a)^2]$, where $\delta_c$ is the center deflection (equilibrium position), $a$ is the radius of the pixel, and $r$ is the radial distance away from the pixel’s center \cite{Santiago2018Graphene}. The refraction index of monolayer graphene can be expressed as $n_g=3.0+i(C_1/3)\lambda$, where $C_1=5.446\mu m^{-1}$, $\lambda$ is the wavelength, and the thickness of the monolayer graphene is 0.34 nm \cite{2018Broad}. The distances between the graphene pixel and the Bragg mirrors are $D_1$ and $D_2$, respectively, as shown in Fig. \ref{Fig1}(a). 

Because the graphene is approximately parallel to the Bragg mirror, and the size of the graphene is much larger than the distance between the up and low Bragg mirrors. When a voltage is applied between the graphene and the conductive layer in the Bragg mirror below, the graphene deflects, but only slightly. Thus, the graphene can be divided into many small regions, each of which is approximately parallel to the layers, and the average transmittance of the structure can be calculated using the transfer matrix method \cite{2018Broad,2019Two}, as detailed in the Appendix A.

Assuming standard illumination (CIE) D65 normally illuminate (white daylight) the proposed structure from the top, the generated colors can be characterized by the CIE XYZ color space with the tri-stimulus values given as \cite{doi:10.1021/acsnano.9b07523,doi:10.1021/acsnano.6b08465,doi:https://doi.org/10.1002/0470020326.ch4,2020Symmetric}.
\begin{equation}
\begin{cases}
X=k\int_{\lambda}E(\lambda)T(\lambda)\overline{x}(\lambda)d\lambda\\
Y=k\int_{\lambda}E(\lambda)T(\lambda)\overline{y}(\lambda)d\lambda\\
Z=k\int_{\lambda}E(\lambda)T(\lambda)\overline{z}(\lambda)d\lambda
\end{cases}\label{eq4}\\
\end{equation}
where $k=100/\int_{\lambda}E(\lambda)\overline{y}(\lambda)d\lambda$, and it is a normalizing factor, $\lambda$ is the wavelength, $E(\lambda)$ is the relative spectral energy distribution of an illuminant, $\overline{x}(\lambda)$, $\overline{y}(\lambda)$ and $\overline{z}(\lambda)$ are the color matching functions for the CIE 1931 $2^\circ$ standard observer, $T(\lambda)$ is the spectral transmittance spectrum.

The chromaticity coordinates can be computed from the tri-stimulus values by \cite{doi:https://doi.org/10.1002/0470020326.ch4}.
\begin{equation}
\begin{cases}
x=\frac{X}{X+Y+Z}\\
y=\frac{Y}{X+Y+Z}
\end{cases}\label{eq5}\\
\end{equation}

Each point (x,y) on the diagram represents a color described by its tristimulus value (X,Y,Z). In the CIE1931 chromaticity space, it is convenient to visualize the color trends \cite{2019Colors}.

\section{Numerical result analysis}

\subsection{Optical response of the microcavity}

\begin{figure}
	\centering
	\includegraphics[width=\linewidth]{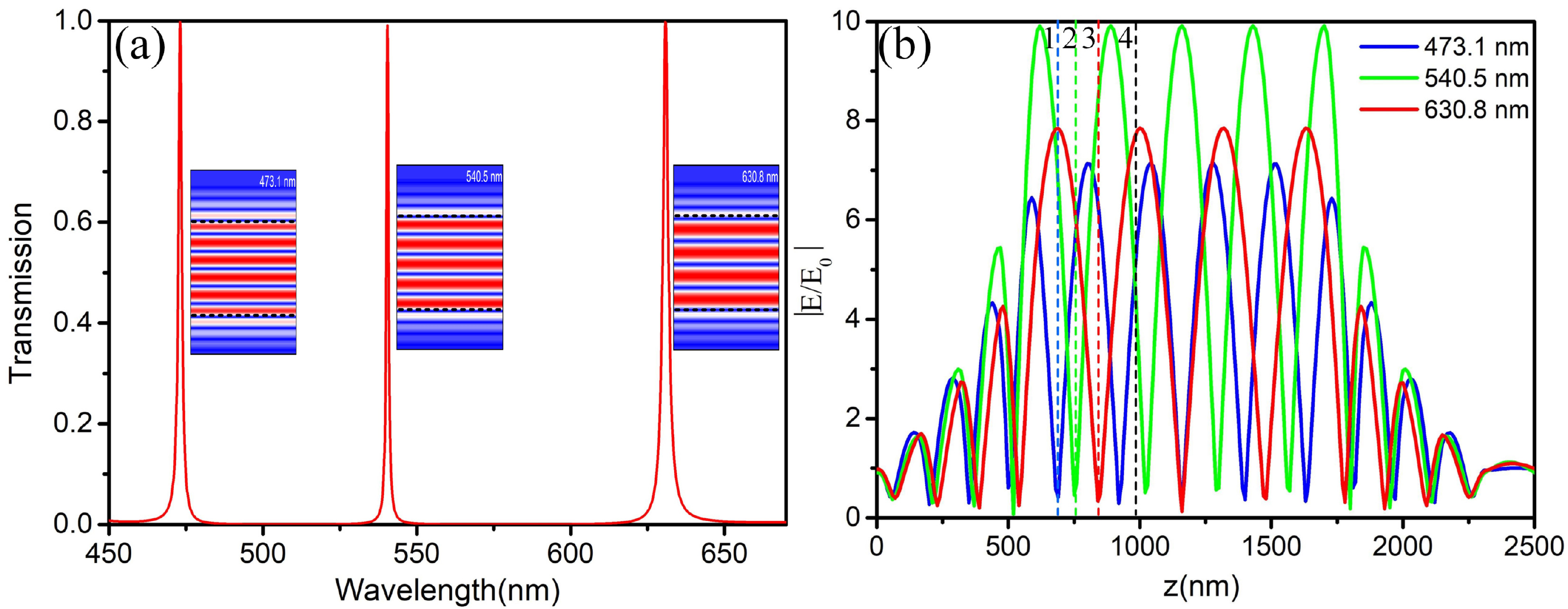}
	\caption{(a)Transmittance of the structure (The insets show the distribution of electric fields in the three modes, and the black dotted areas in the insets is air cavity). (b)Electric field intensity distribution at the resonant wavelength of the three modes (dashed lines 1-4 are the corresponding positions of 0, 70, 170 and 320 nm in the microcavity in Fig. \ref{Fig3}).}
	\label{Fig2}
\end{figure}

Light can be coupled into the optical microcavity from the normal direction. The resonant state of the microcavity is sensitive to the thickness of the intermediate air gap layer, and the resonant wavelength $\lambda_c$ satisfies the following condition $m_i\lambda_c/2\propto L_c$, where $L_c$ represents the optical path of the microcavity, and $m_i$ is a positive integer \cite{2018Broad}. When the microcavity length is set as shown in Fig. \ref{Fig1}(a), there are three resonance modes in the microcavity, the resonance peaks of which appear at the wavelengths of 473.1 nm, 540.5 nm and 630.8 nm, respectively. The transmission spectrum of the microcavity is shown in Fig. \ref{Fig2}(a). The localized light fields are formed in three modes at different positions of the microcavity, so there are three resonant transmission peaks in the spectral region, and the transimittance of which are 99.9\%, 99.1\% and 99.8\%, respectively. It can be seen that the transmission spectrum shows the R, G and B modes necessary to create white. The generated resonances are narrow with a full width at half-maximum ranging between 0.8 and 2.1 nm. Thus, the quality factors Q of these three transmission peaks are 394.3, 675.6 and 300.4, respectively. The higher Q values ensure the generation of a color with higher saturation. The electric field distribution diagram as shown in Fig. \ref{Fig2}(b) is enhanced and and its electric energy ratio $|E/E_0|$ is greater than 7 in the three modes, here $E$ and $E_0$ represent the amplitude of local electric field and the amplitude of incident photoelectric field respectively. In calculation, we usually set the value of $E_0$ as 1. The reason is that the incident light interferes with the reflected light of the Bragg mirror generate three groups of light standing waves. As can be seen from the insets in Fig. \ref{Fig2}(a), when the resonant wavelength is 473.1 nm, 540.5 nm and 630.8 nm, the electric field at the resonant wavelength of the three modes is effectively enhanced to produce a standing wave pattern. 

\subsection{Color modulation using graphene with different deflection}

According to the three modes of R, G and B presented by the photonic crystal microcavity transmission spectrum, if the relative intensities of the three modes are adjusted according to the human eye's perception for these three colors, the colors of CIE gamut can be continuously covered \cite{doi:10.1021/acsnano.9b07523}. When we coupled five-layer graphene mechanical pixels with a diameter of 10 $\mu$m into the above three-mode photonic crystal microcavity, the microcavity greatly enhanced the interaction between light and graphene. Meanwhile, the intensity of the transmission spectrum of the three modes in the microcavity can be governed by graphene deflection. The mechanism is the modulation of light absorption caused by graphene mechanical deflection. In other words, when graphene is located  in the abdomen of the standing wave (with a strong light field), it exhibits a stronger absorption and therefore the transmittance of the structure is lower. On the contrary, when graphene is located in the node of the standing wave (with a weak light field), it has a weaker absorption and therefore a higher transmittance of the structure. In this way, when graphene located between the standing wave node and the wave abdomen, the transmitted light intensity changes harmoniously with the change of graphene's equilibrium position in the cavity. Due to the different field distribution of the three resonance modes, the absorption of light with different colors can be adjusted by tuning the equilibrium position of graphene in the microcavity, allowing the single red, green and blue color or their mixed colors to pass through. Additionally, the equilibrium position of graphene membranes can be controlled by voltage \cite{Santiago2018Graphene}, that is, the modulation of both the color and the intensity of the three-mode transmission can be tuned dynamically by voltage.
\begin{figure}
	\centering
	\includegraphics[width=\linewidth]{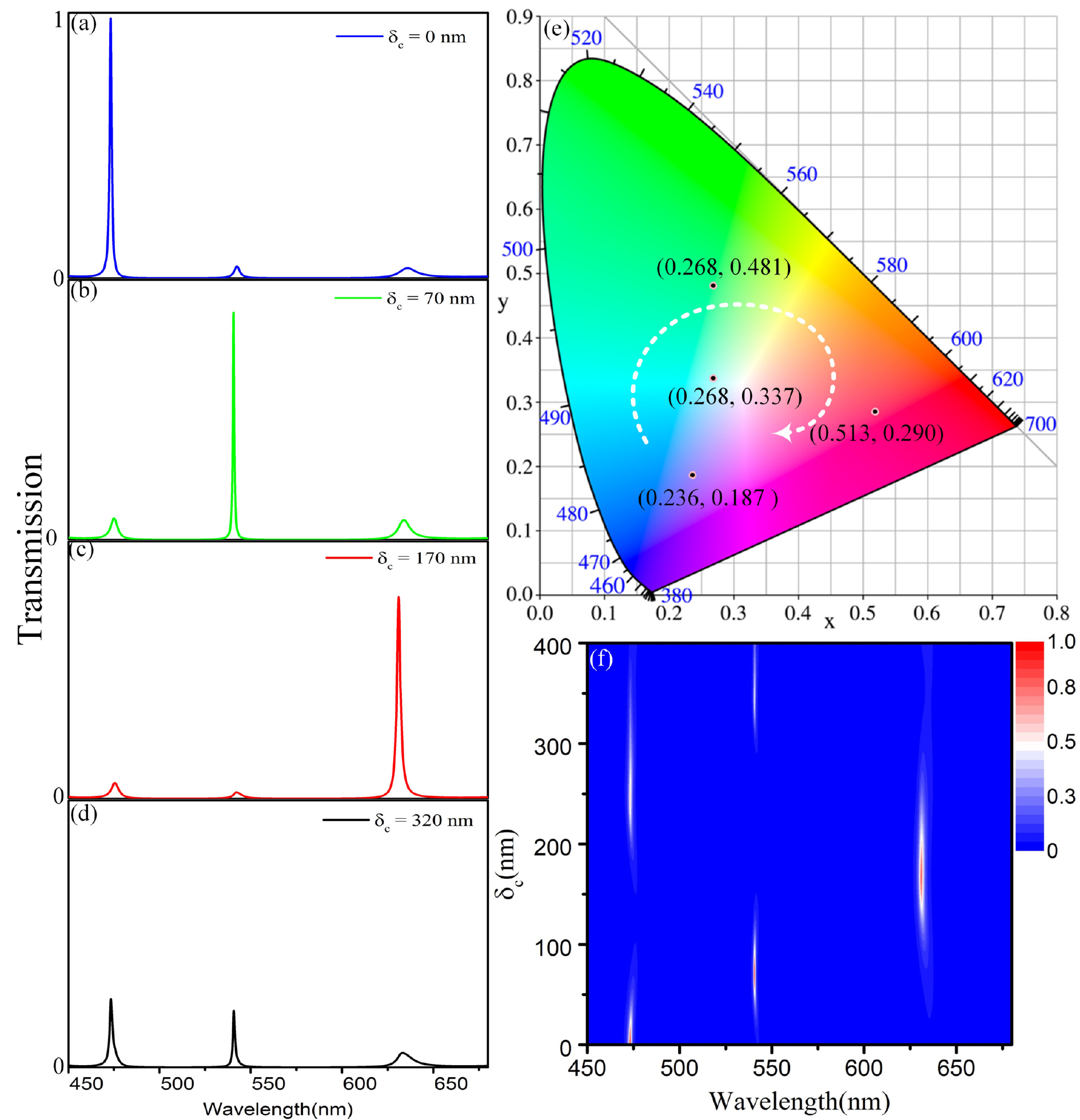}
	\caption{(a)-(d) Transmission spectrum of the system when the equilibrium position of graphene is 0, 70, 170, 320 nm, respectively. (e) The color coordinates of 0, 70, 170, 320 nm and the corresponding positions in 1931 CIE chromaticity diagram. The white dashed lines with the arrow show the color trend as the deflection of graphene increases. (f) Simulation of the averaged transmission as a function of the equilibrium position of graphene.}
	\label{Fig3}
\end{figure}

As a typical example, four extraction equilibrium positions of graphene are: 0 nm, 70 nm, 170 nm and 320 nm, which are used as the research objects. The corresponding dashed line positions 1-4 in the microcavity are present in Fig. \ref{Fig2}(b), and the corresponding transmission spectra are shown in Fig.  \ref{Fig3}(a)-(d), respectively. When the equilibrium position of graphene is 0 nm, graphene is located at the wave node of B mode, and it displays a lower absorption to B mode and a weaker influence on the electric field distribution of B mode, while a higher absorption to G and R mode. Therefore, the graphene membrane allows B mode to pass through, where the transmission peak of B mode is the highest and the interference color appears in blue, as shown in Fig. \ref{Fig3}(a). Similarly, when the equilibrium position is 70 nm (170 nm), graphene located at the wave node of G (R) mode exhibits a lower absorption to G (R) mode and a weaker influence on the electric field distribution of G (R) mode, the interference color is green (red), as shown in Fig. \ref{Fig3}(b)-(c). When the transmittance of one mode reaches the maximum, the other two modes are lower and close to zero, and the color formed is of a non-monochromatic light, but the color at these points reveals a higher saturation. As described in Fig. \ref{Fig3}(e), the color coordinates of the above three cases are located in (0.236, 0.187), (0.268, 0.481) and (0.513, 0.290), respectively, which fully reflects the wide gamut of color modulation. When the equilibrium position of graphene is 320 nm, the graphene membrane is located near the wave abdomen of the three standing waves. Affected by the strong light fields of the three modes, the absorption of graphene to the three modes is high, thus the transmittance of the three modes is low, as shown in Fig. \ref{Fig3}(d). Its interference color is close to white as depicted in the CIE chromaticity diagram, and its color coordinate is (0.268, 0.337), marked in Fig. \ref{Fig3}(e). As shown in Fig. \ref{Fig3}(f), the average transmittance of the microcavity changes with the equilibrium position of graphene, indicated that the graphene membranes with different deflections affect directly the light field distribution of the whole device when the three standing waves are propagated in the microcavity. In fact, there is more than one place where graphene can achieve the above effect. Through analysis, we can also find other placement positions that achieve the similar conclusions, such as the position close to $z$ = 1240 nm in the microcavity (see the Appendix B), which provides a basis for the flexibility of our structural design.

\begin{figure}
	\centering
	\includegraphics[width=\linewidth]{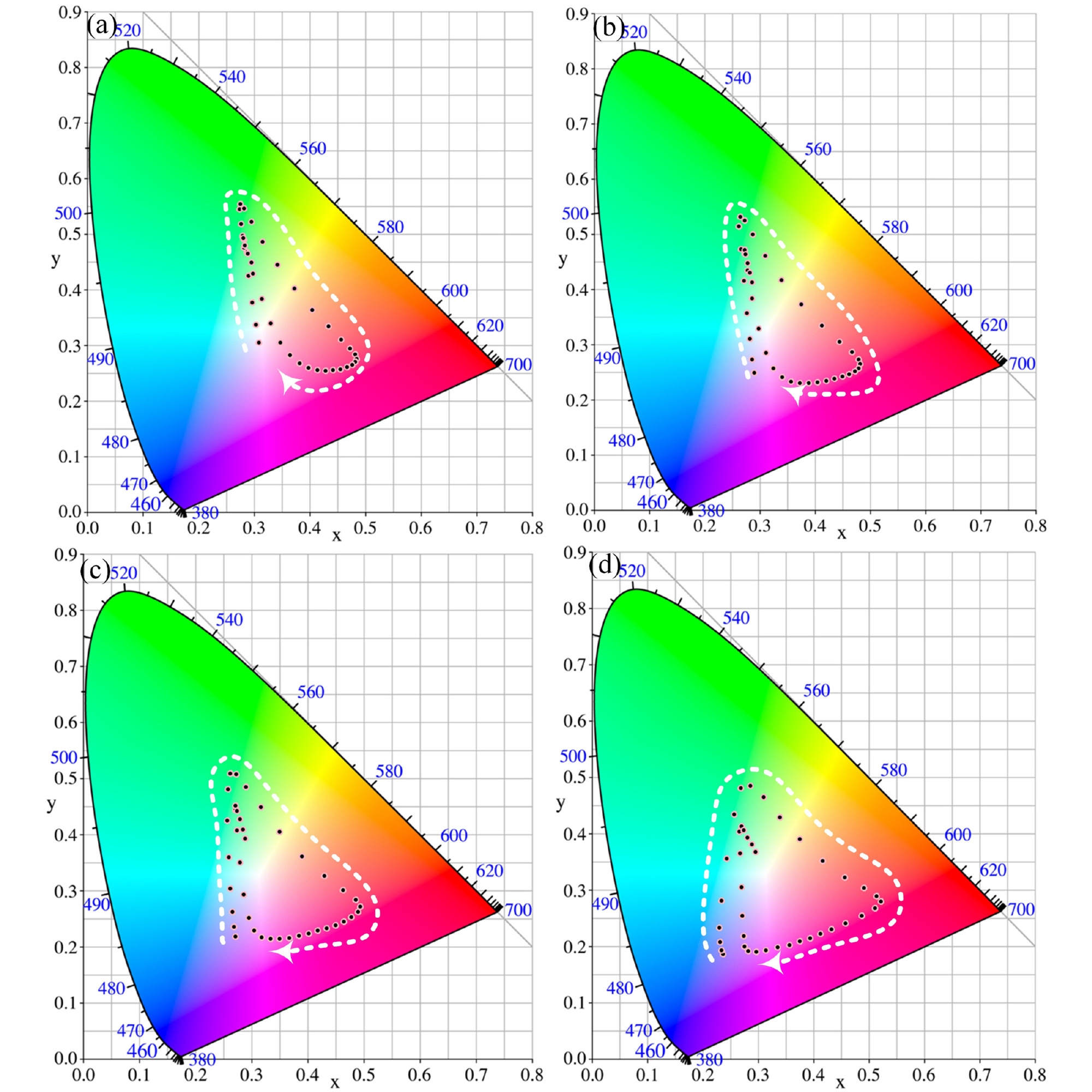}
	\caption{Transmission colors with the number of graphene layers (the white dashed lines with the arrow represent the evolution trend for the colors as the deflection of graphene increases), where (a), (b), (c) and (d) correspond to two-layer, three-layer, four-layer and five-layer of graphene, respectively.}
	\label{Fig4}
\end{figure}

\begin{table*}
	\small
	\caption{Different equilibrium positions of graphene correspond to transmission spectral color coordinates and interference color}
	\label{tbl}
	\begin{tabular*}{\textwidth}{@{\extracolsep{\fill}}m{2cm}<{\centering}|m{3cm}<{\centering} | m{2cm}<{\centering} |  m{2cm}<{\centering}|m{3cm}<{\centering} | m{2cm}<{\centering}}
		\hline
		$\delta_c$(nm) &   CIE(x,y)    &    color   & $\delta_c$(nm) &   CIE(x,y)    &     color \\
		\hline
		0  & (0.236,0.187) & \includegraphics[width=10mm, height=7mm]{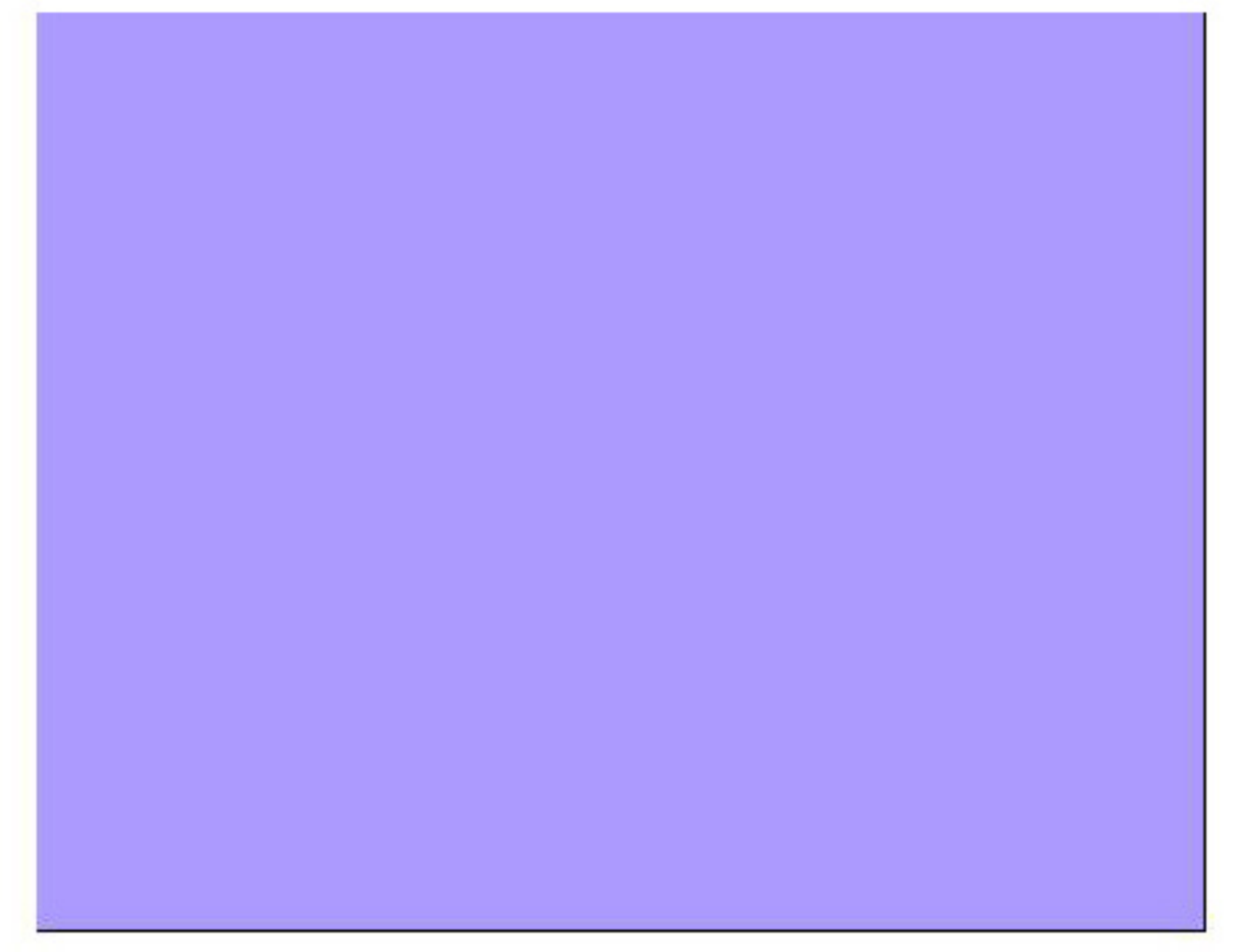}  &   200  & (0.432,0.230) & \includegraphics[width=10mm, height=7mm]{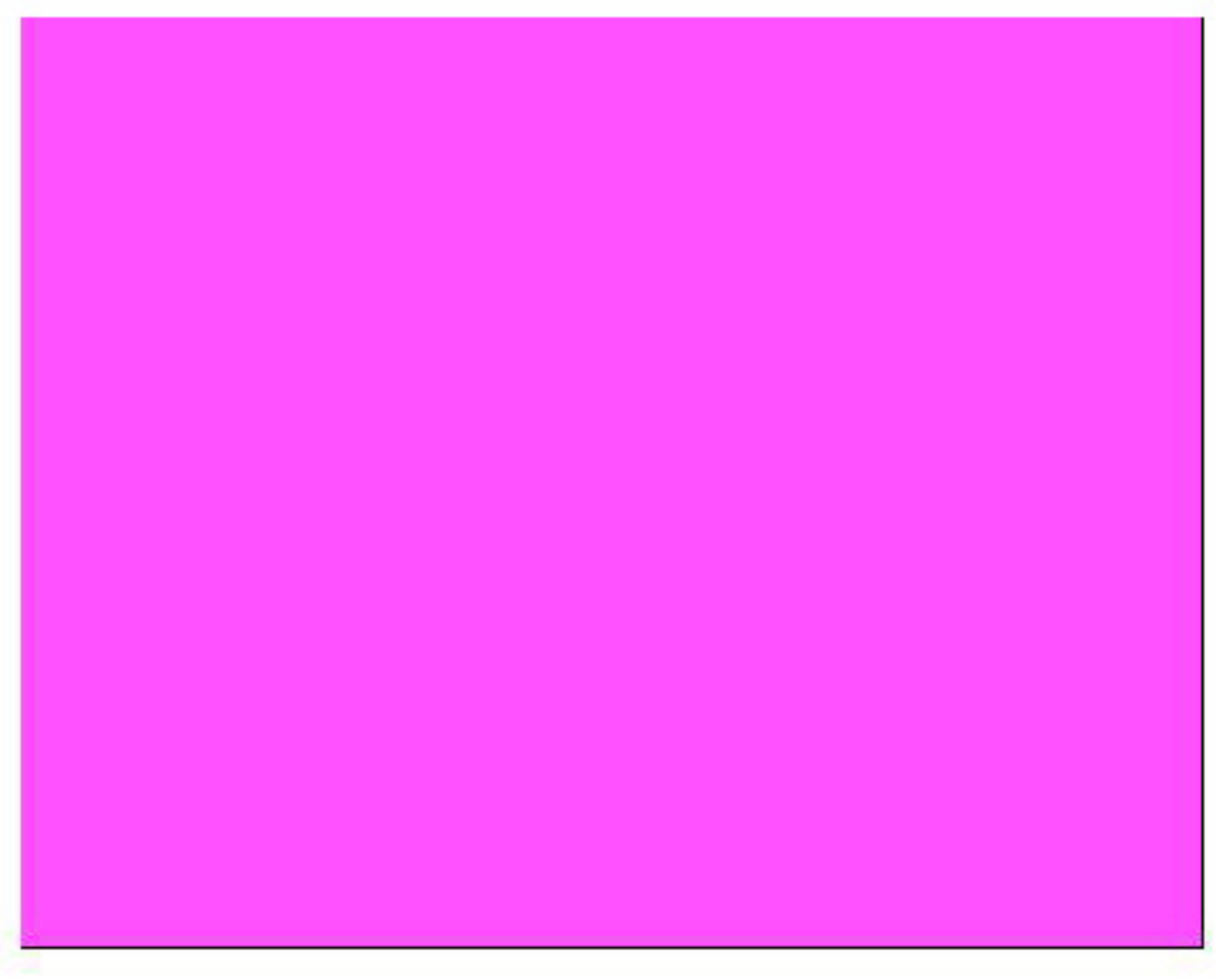} \\
		20  & (0.231,0.208) & \includegraphics[width=10mm, height=7mm]{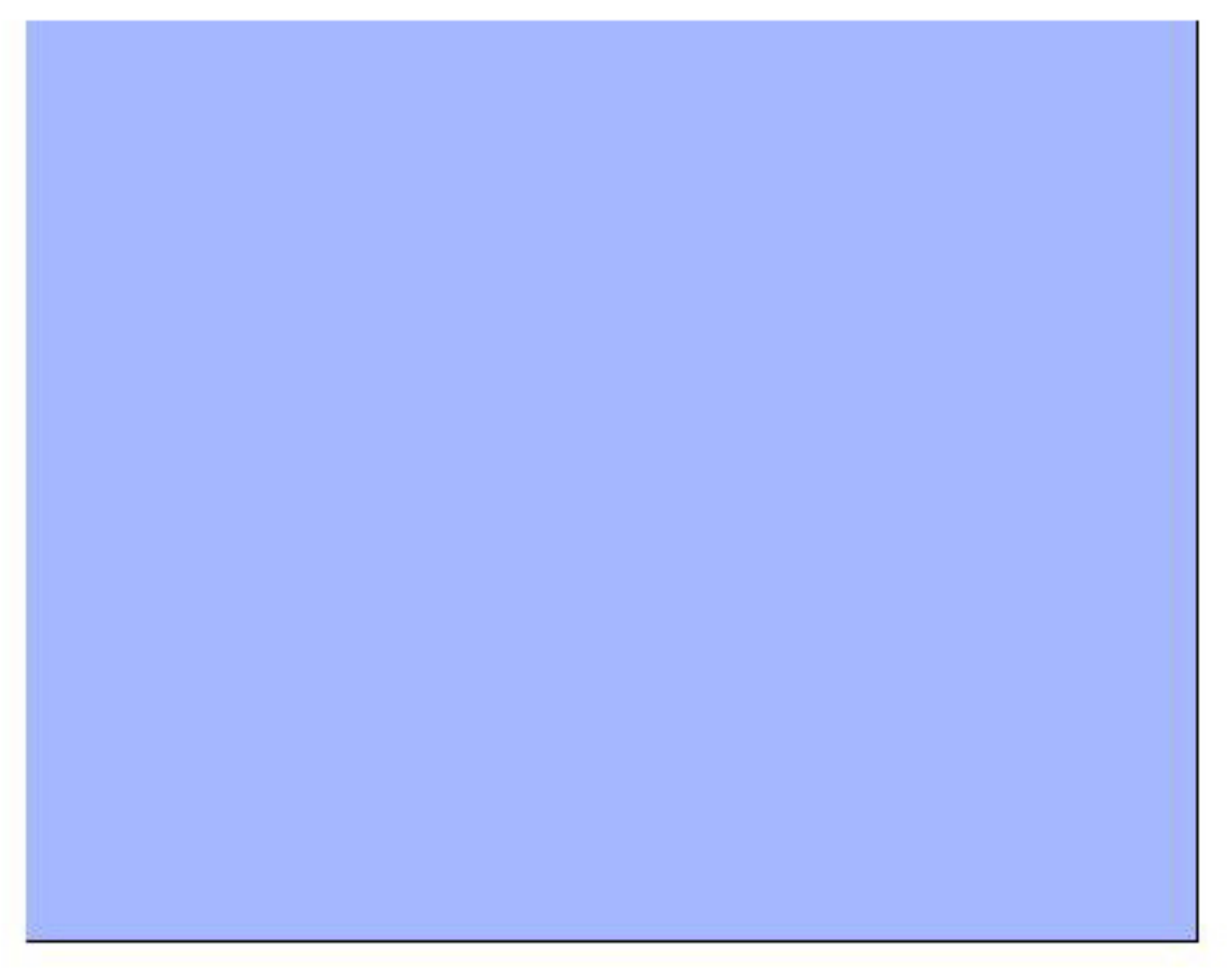} &   220    & (0.394,0.215) & \includegraphics[width=10mm, height=7mm]{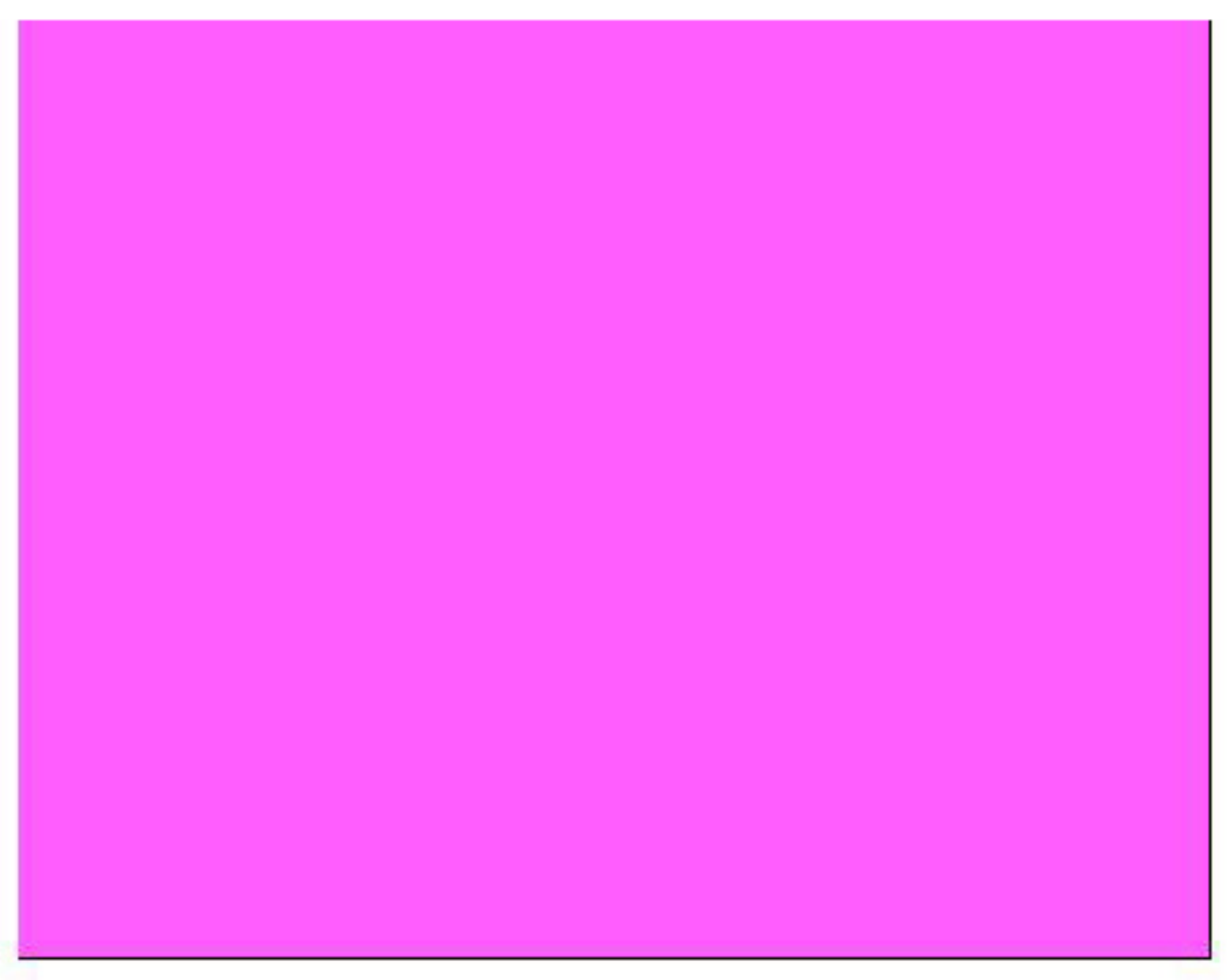}\\
		40  & (0.233,0.281) & \includegraphics[width=10mm, height=7mm]{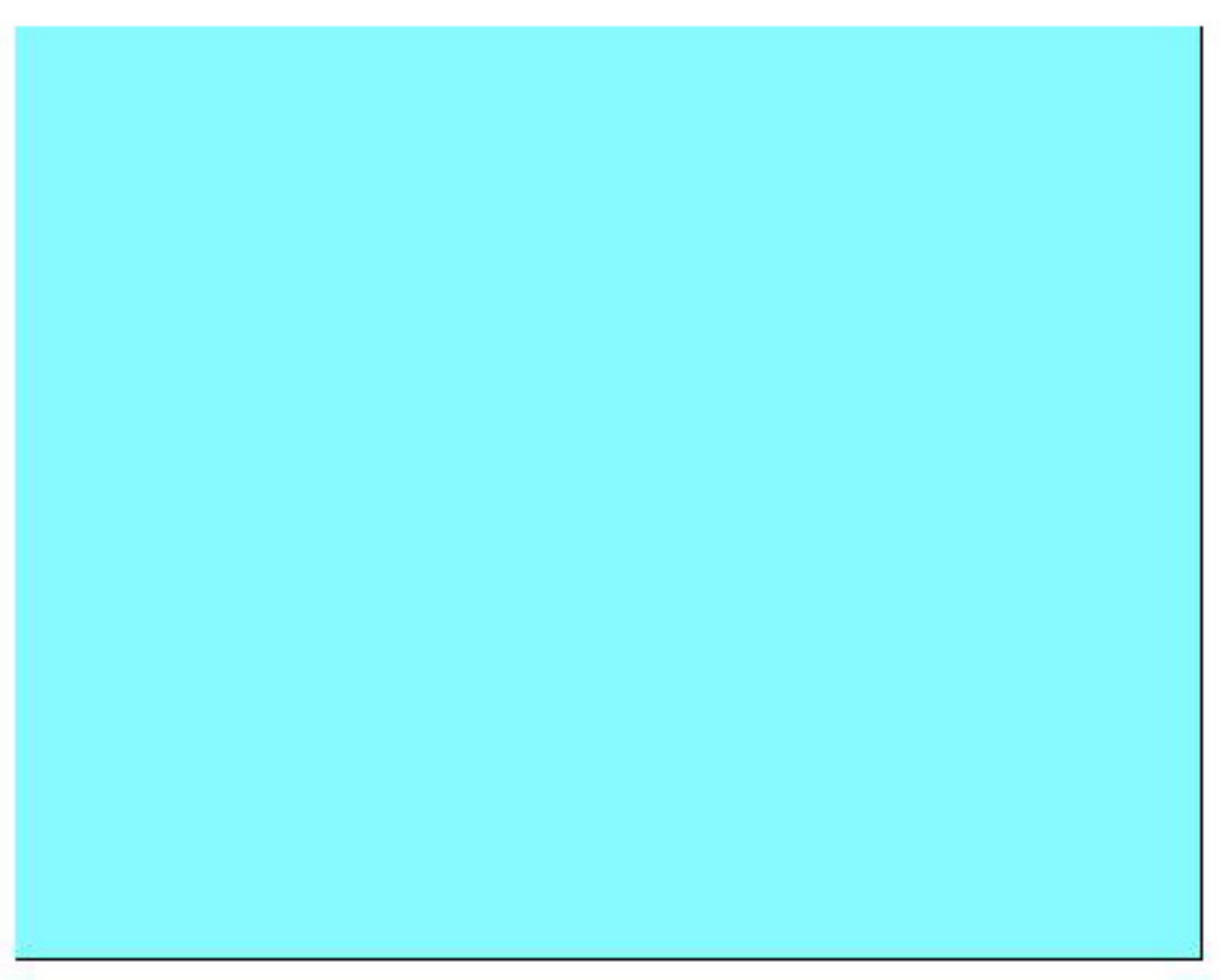} &   240    & (0.356,0.203) & \includegraphics[width=10mm, height=7mm]{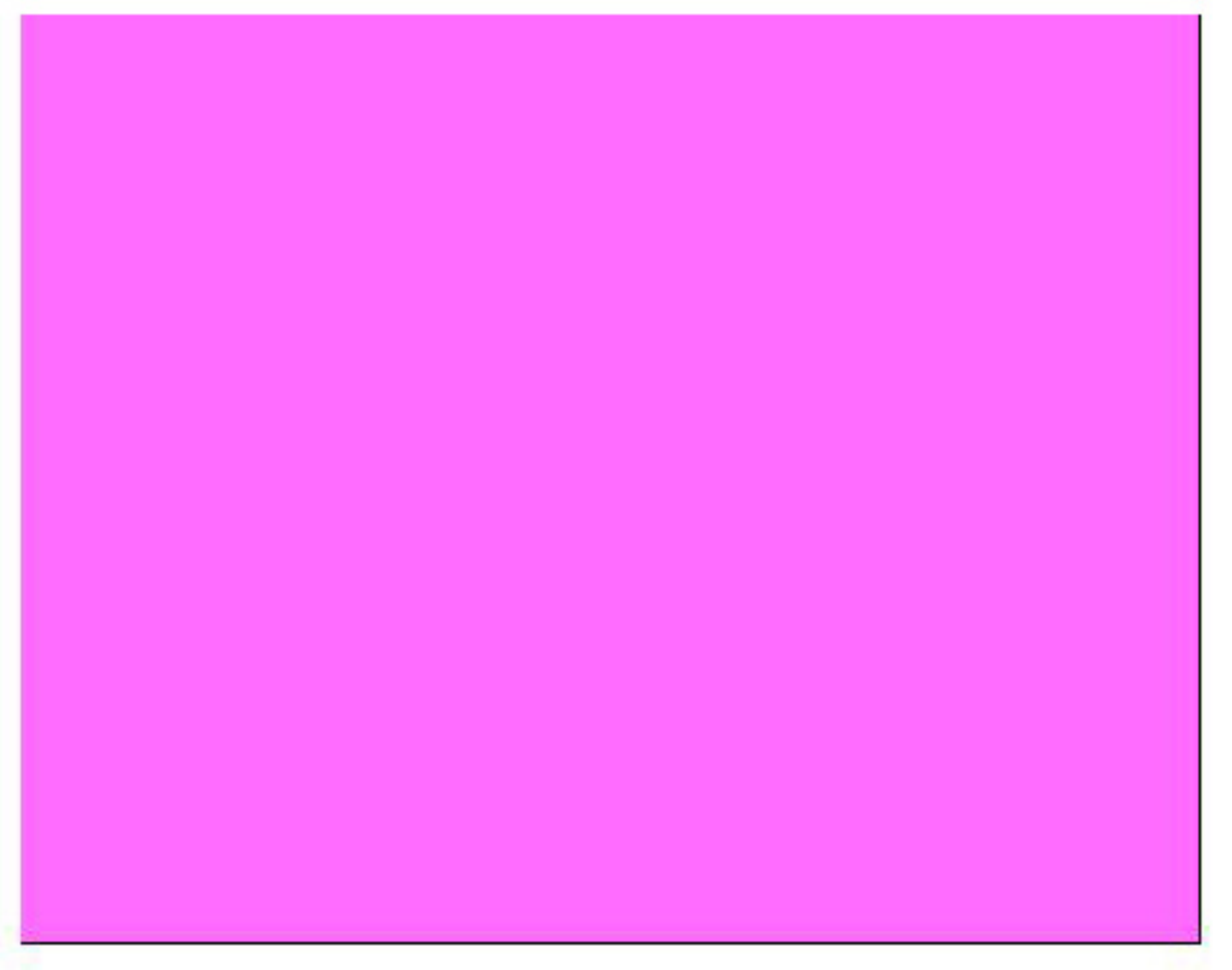}\\
		
		60  & (0.256,0.463) & \includegraphics[width=10mm, height=7mm]{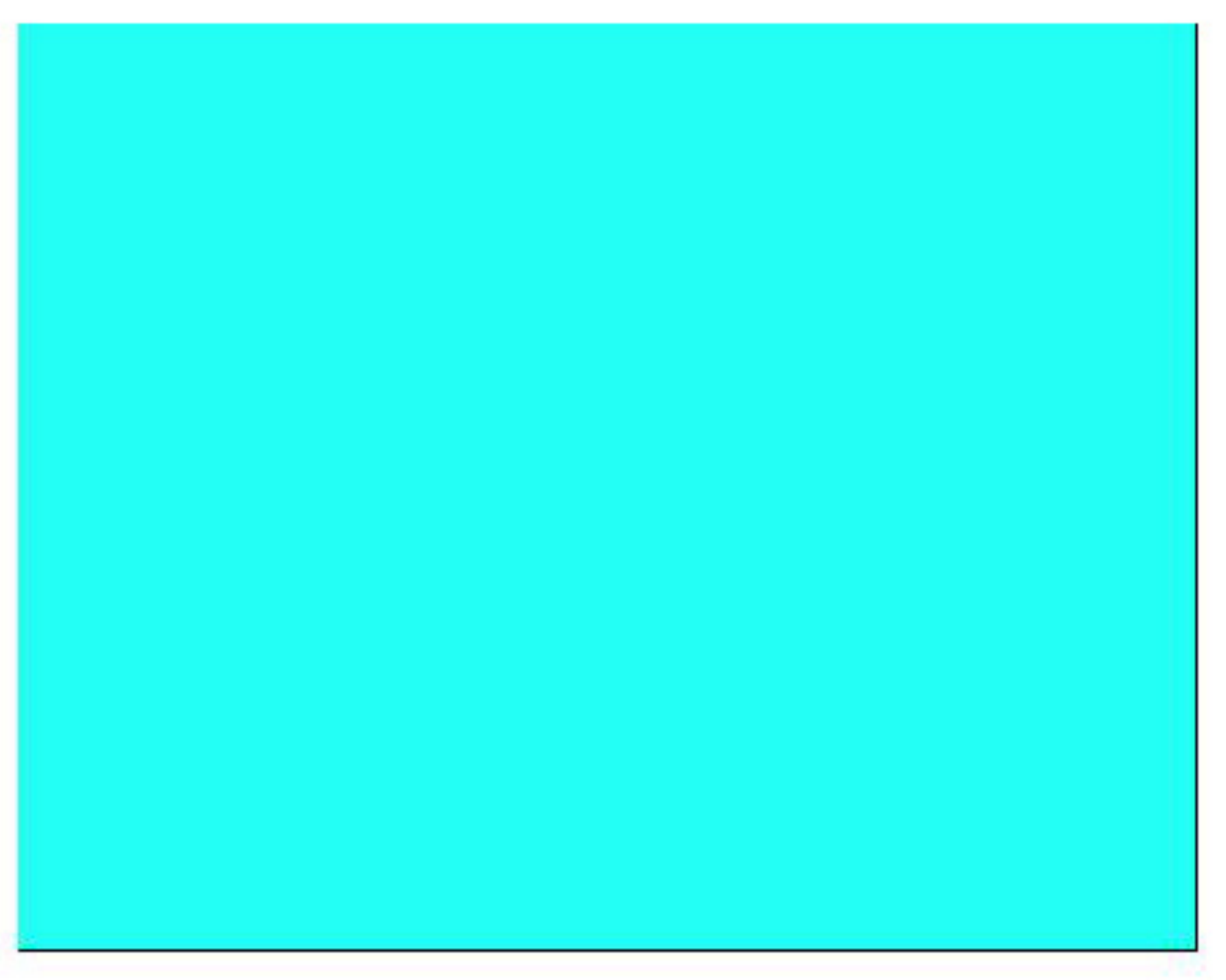} &   260    & (0.314,0.193) & \includegraphics[width=10mm, height=7mm]{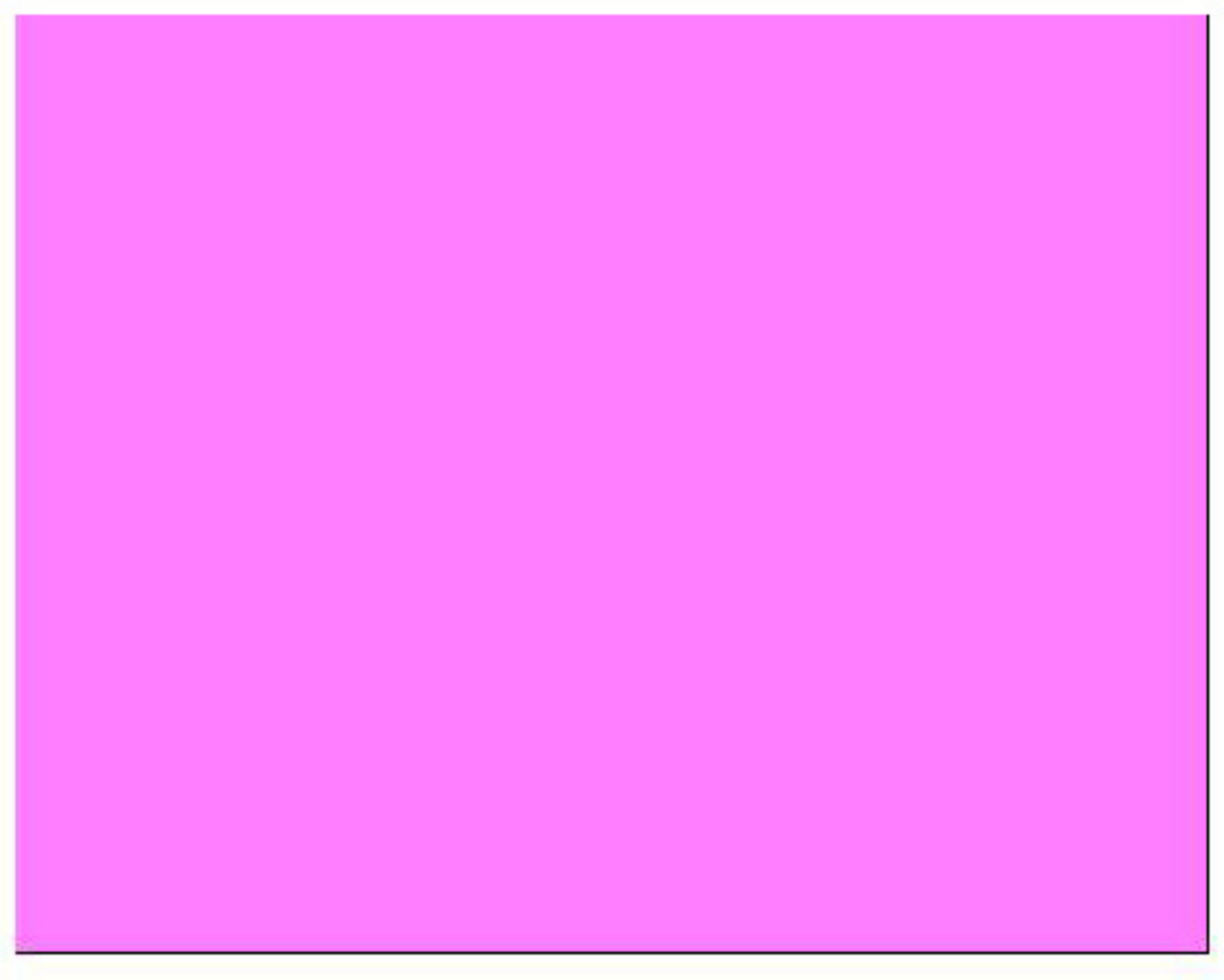}\\
		
		80  & (0.285,0.486) & \includegraphics[width=10mm, height=7mm]{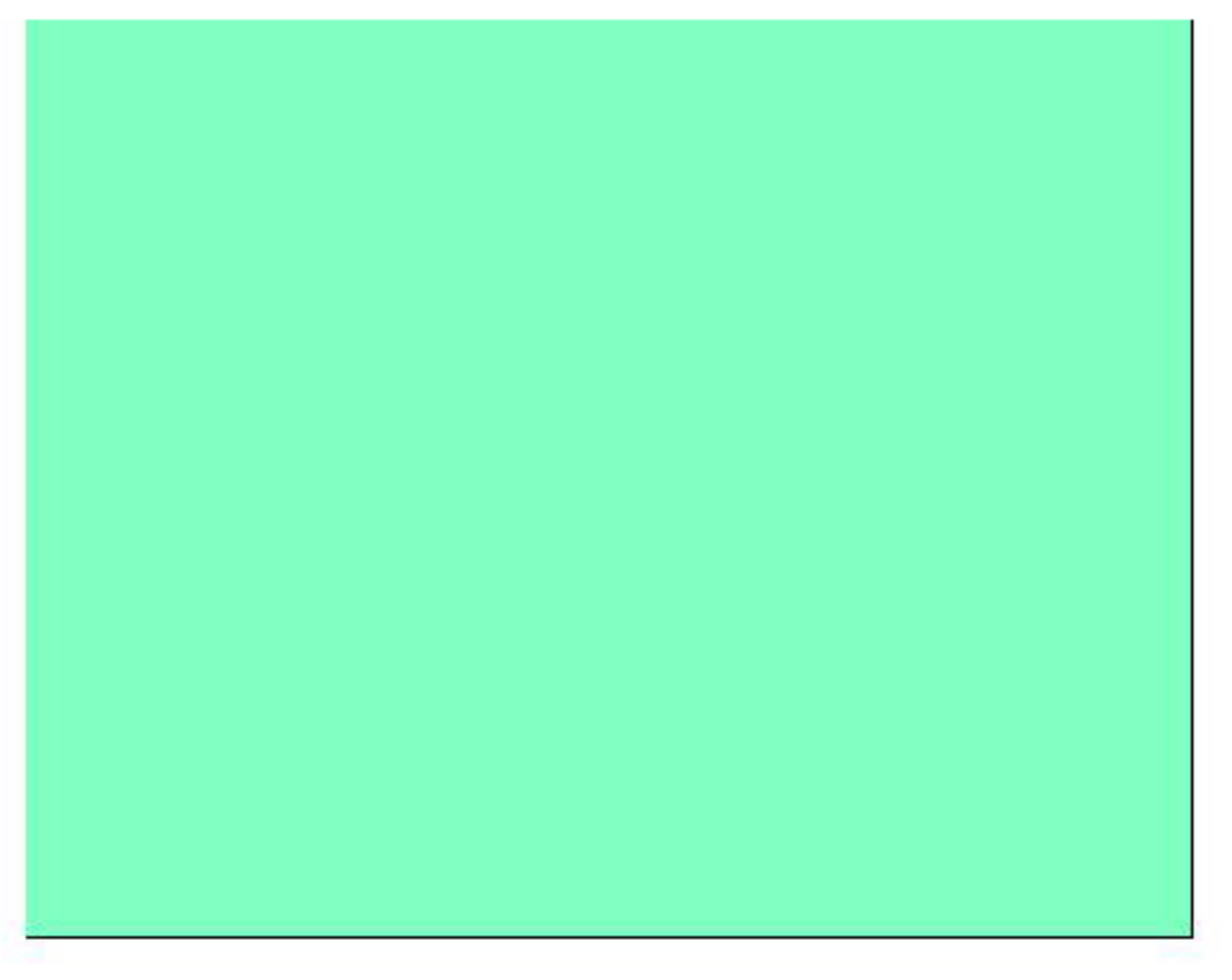} &   280    & (0.282,0.192) & \includegraphics[width=10mm, height=7mm]{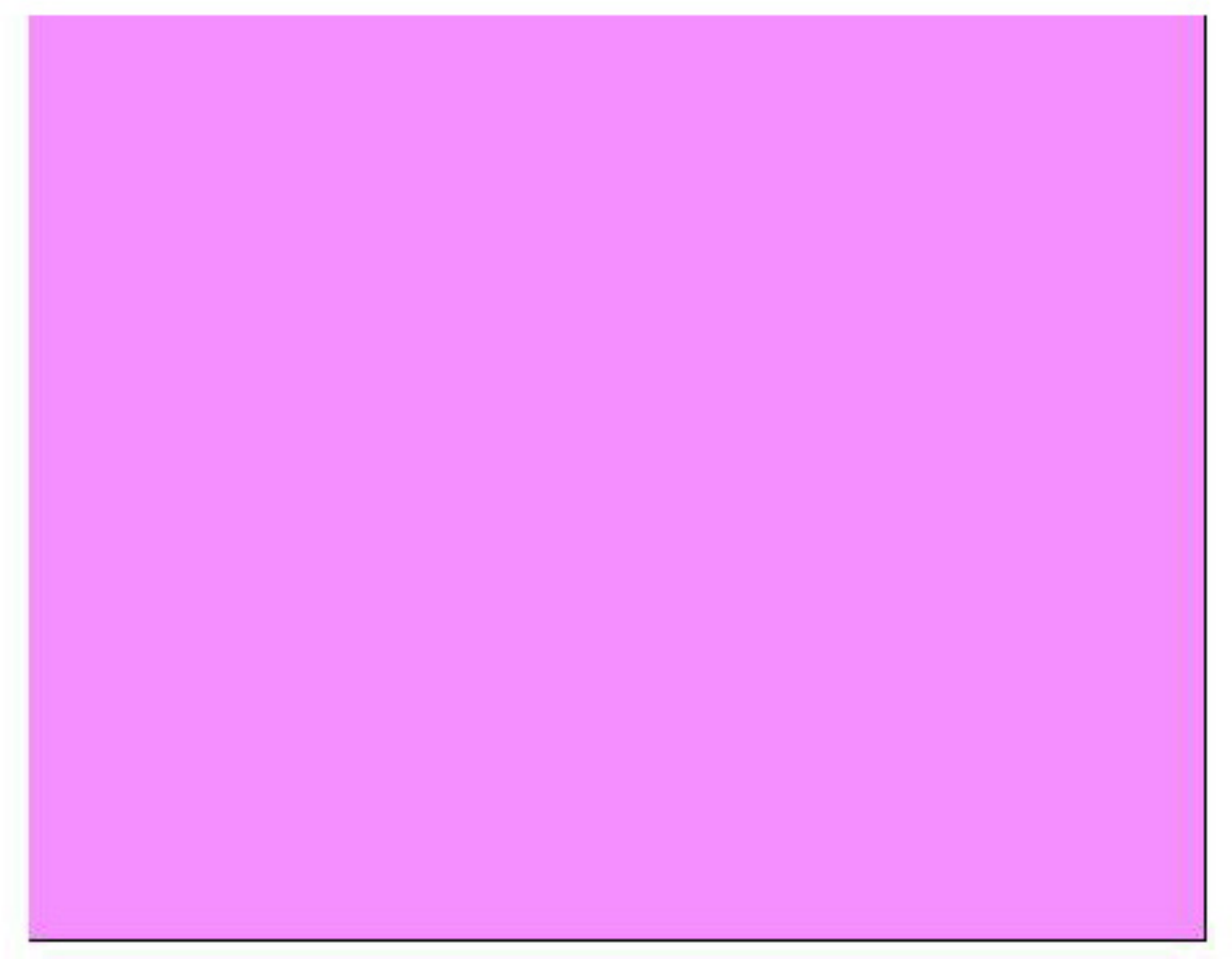}\\
		
		100  & (0.339,0.430) & \includegraphics[width=10mm, height=7mm]{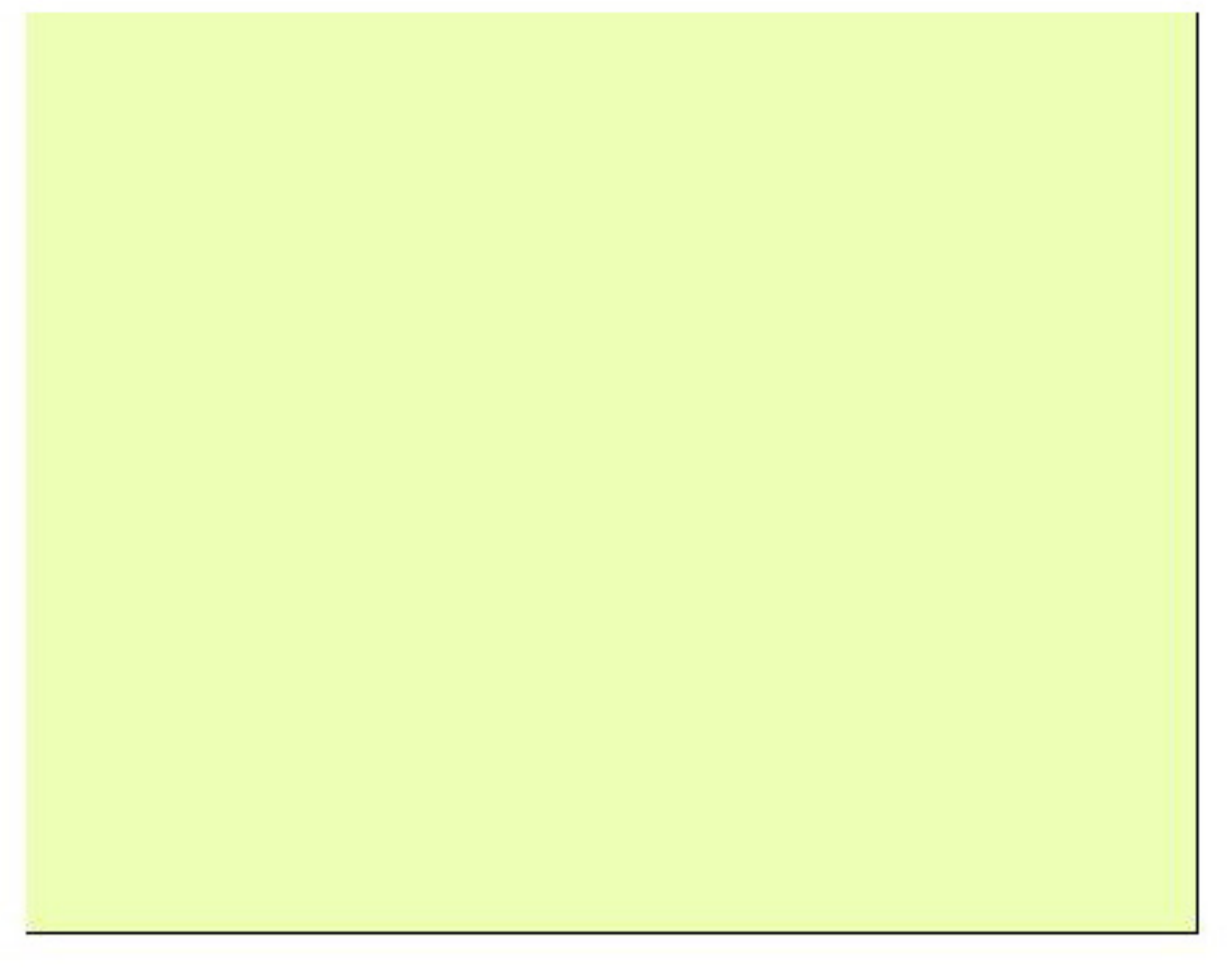} &   300    & (0.273,0.219) & \includegraphics[width=10mm, height=7mm]{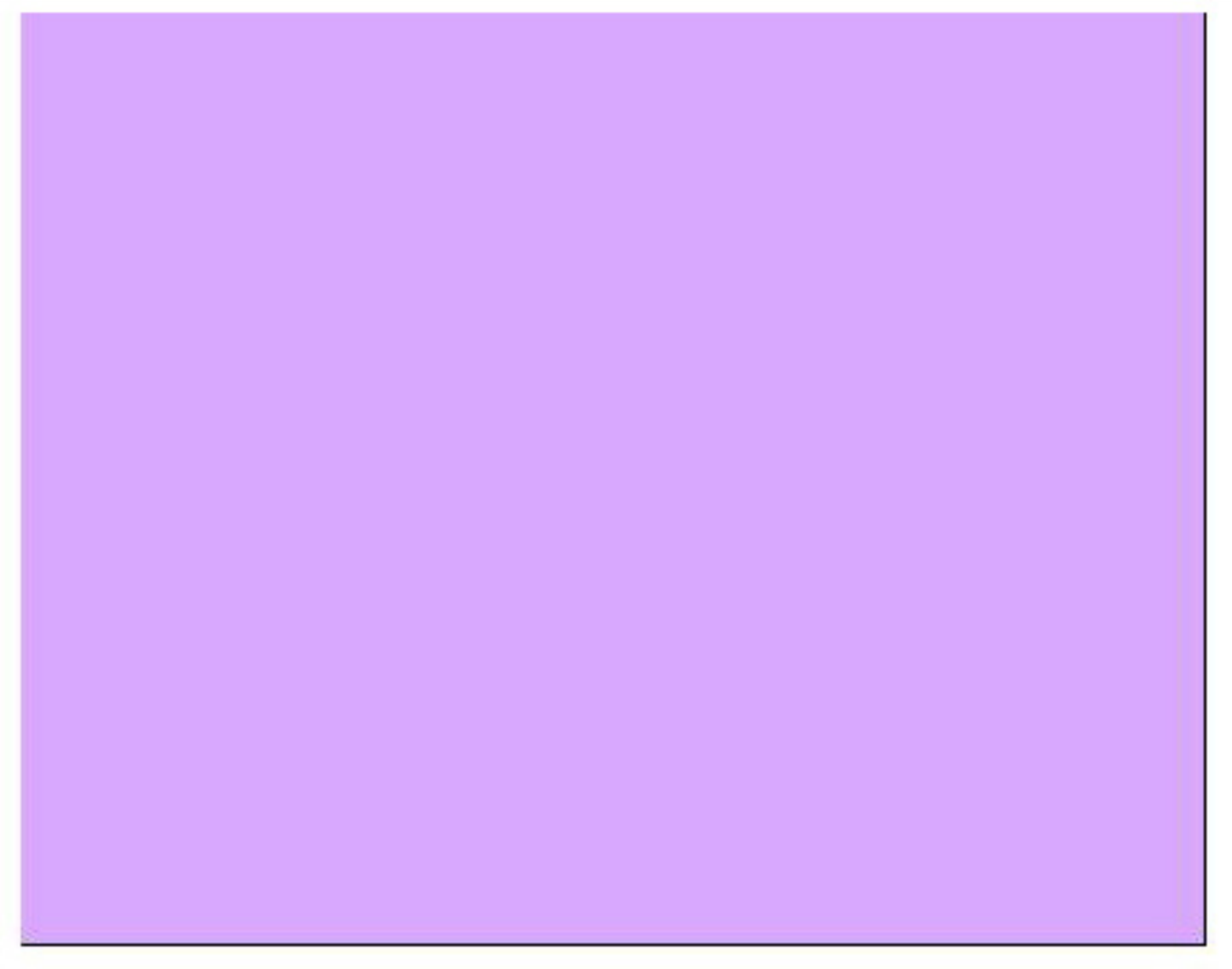}\\    
		
		120  & (0.416,0.352) & \includegraphics[width=10mm, height=7mm]{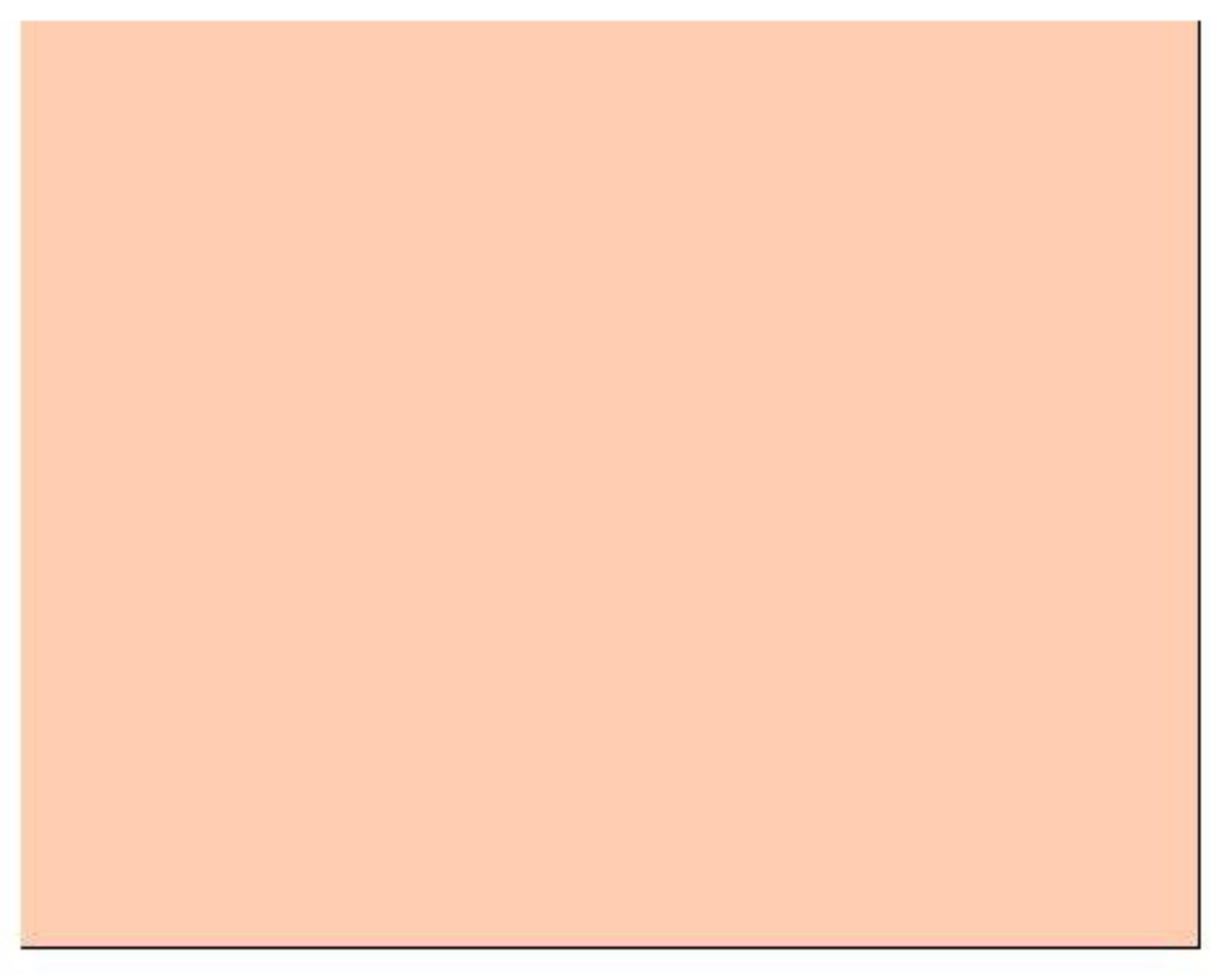} &   320    & (0.268,0.337) & \includegraphics[width=10mm, height=7mm]{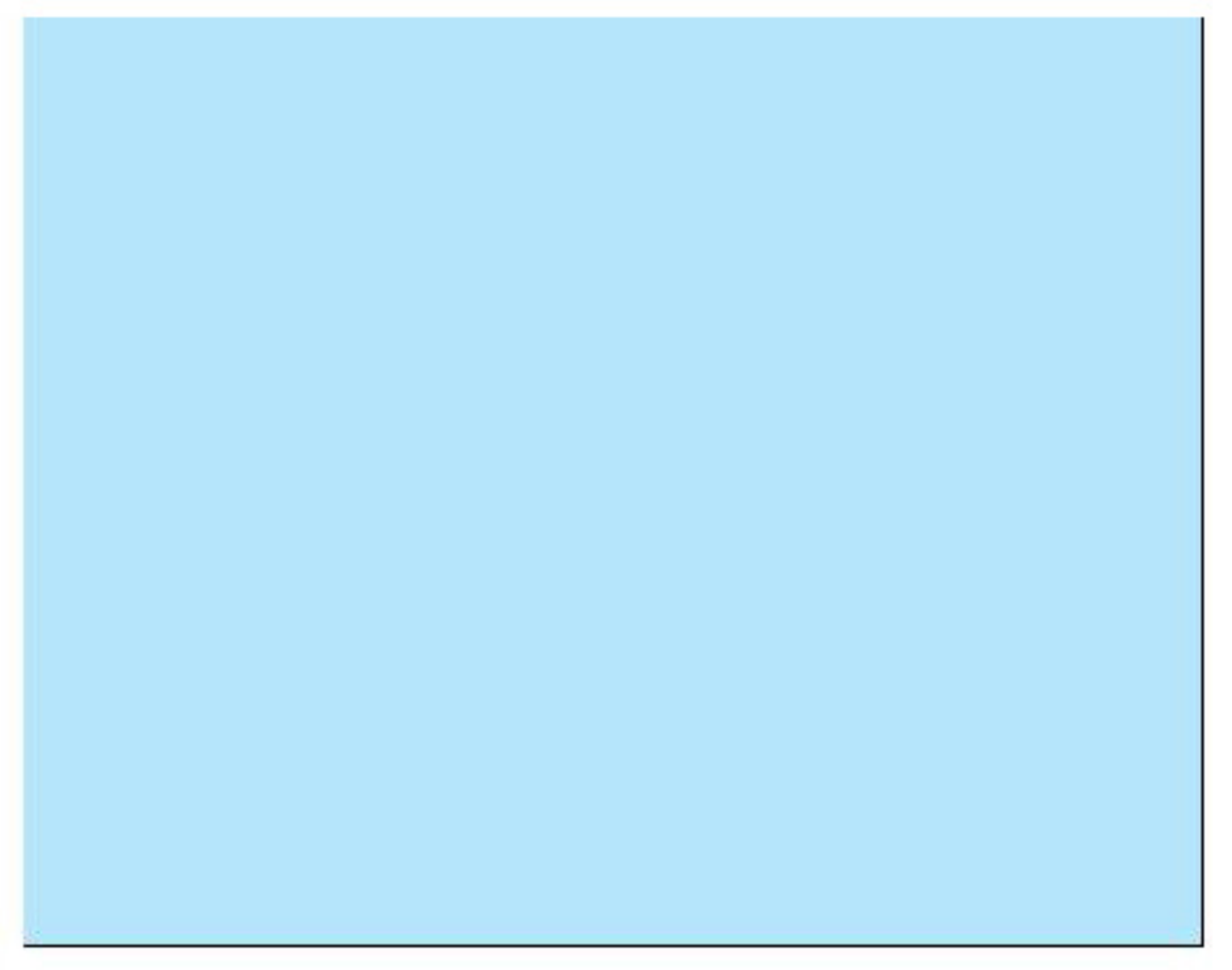}\\    
		
		140  & (0.491,0.304) & \includegraphics[width=10mm, height=7mm]{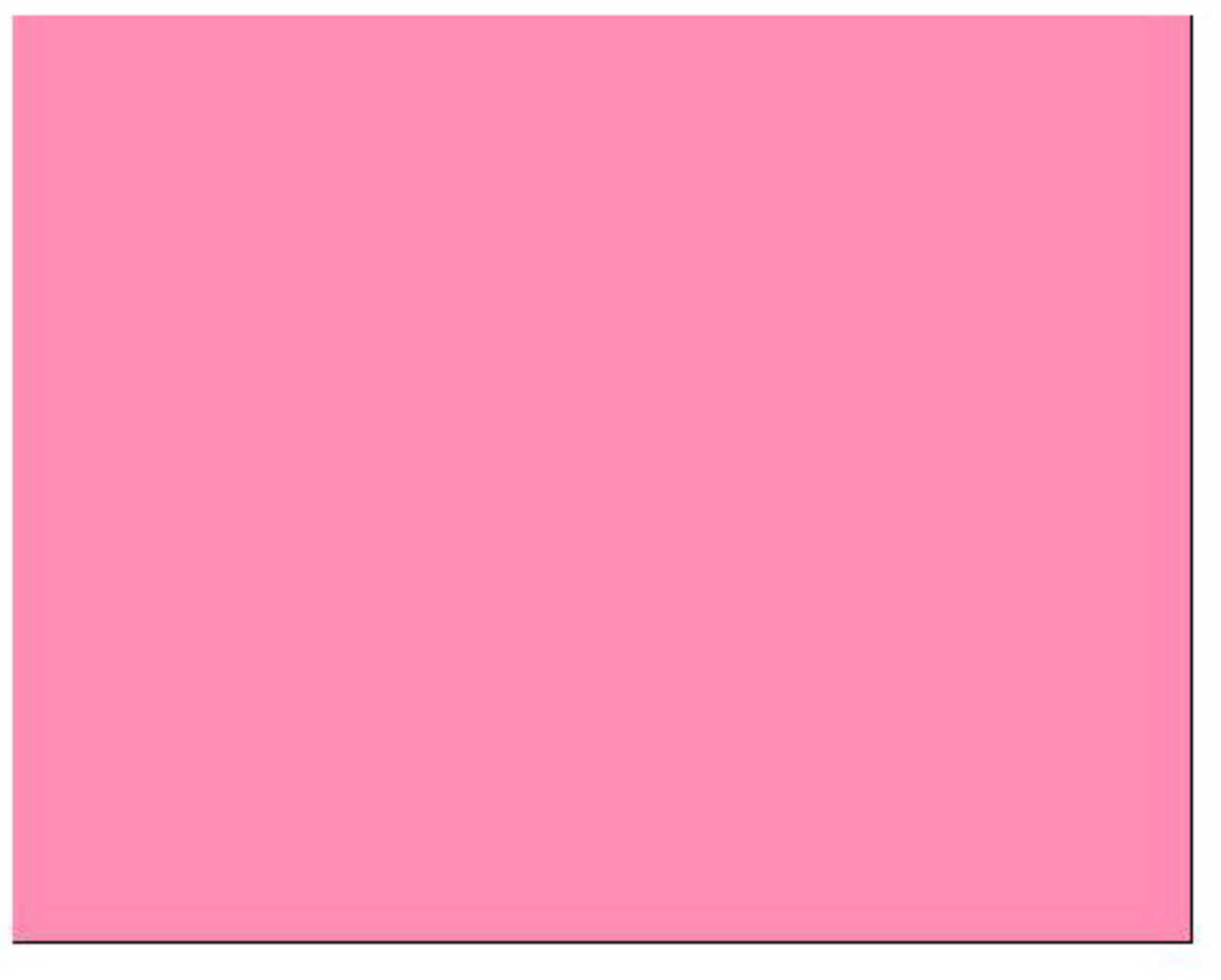} &   340    & (0.266,0.404) & \includegraphics[width=10mm, height=7mm]{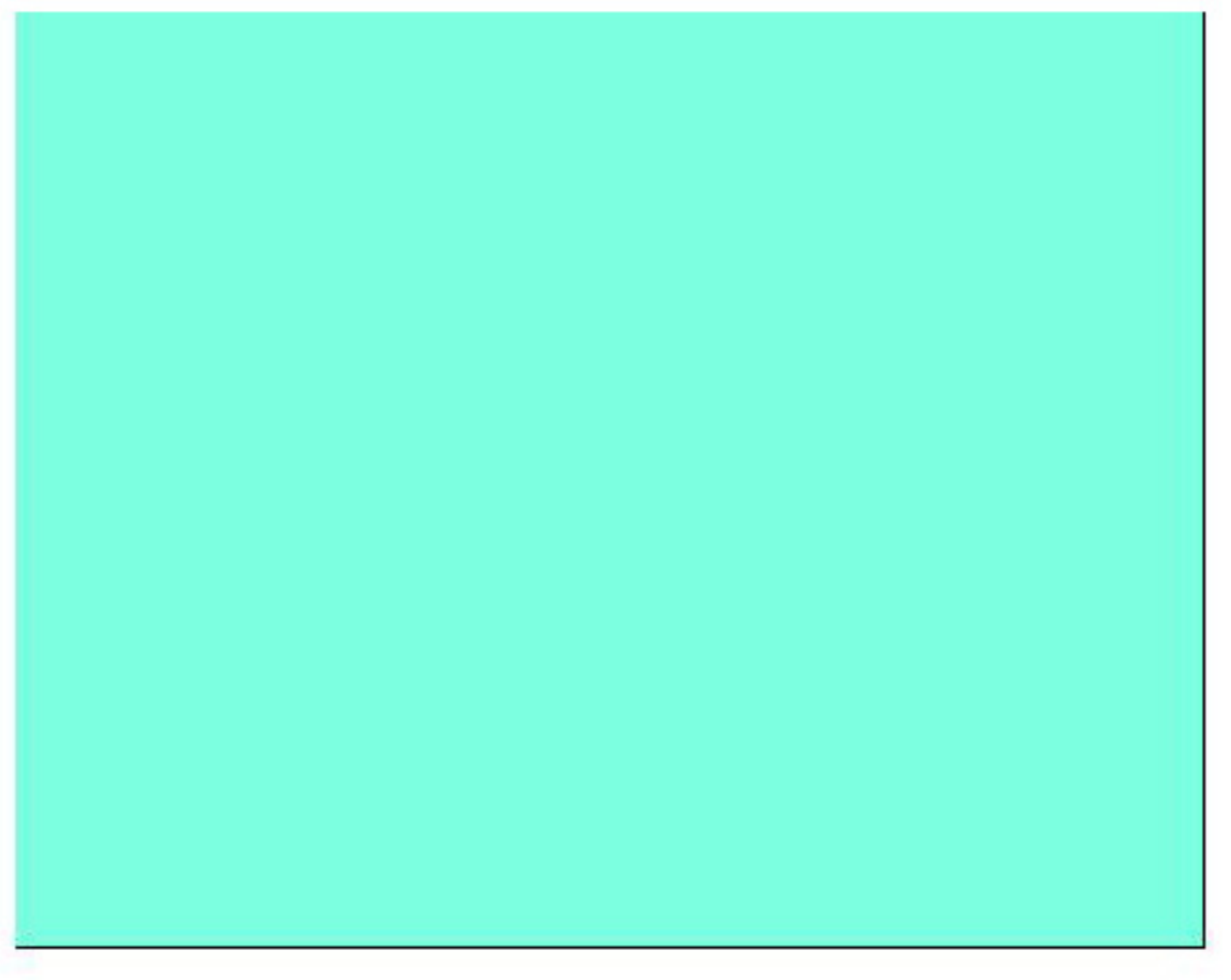}\\  
		
		160  & (0.521,0.280) & \includegraphics[width=10mm, height=7mm]{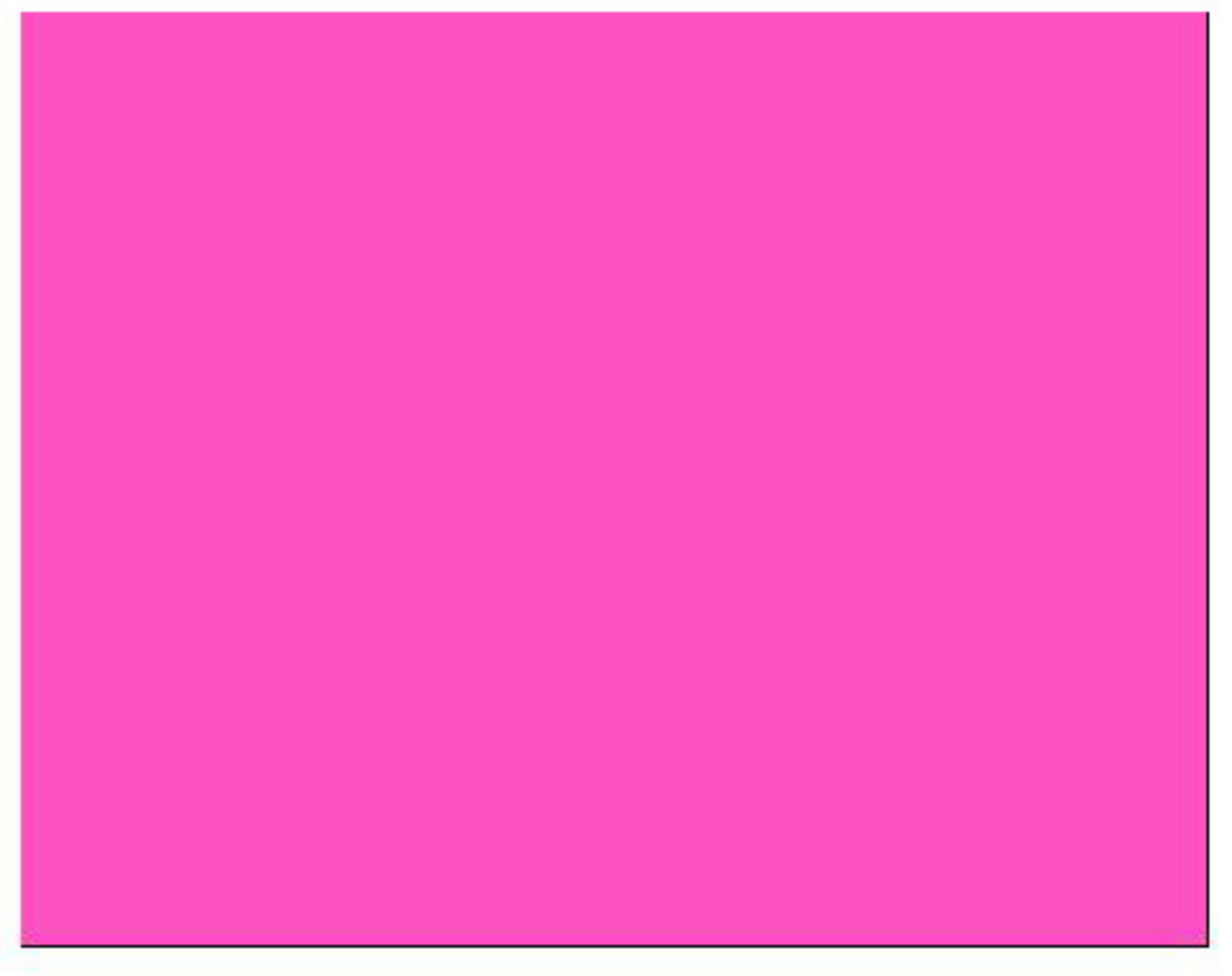} &   360    & (0.274,0.406) & \includegraphics[width=10mm, height=7mm]{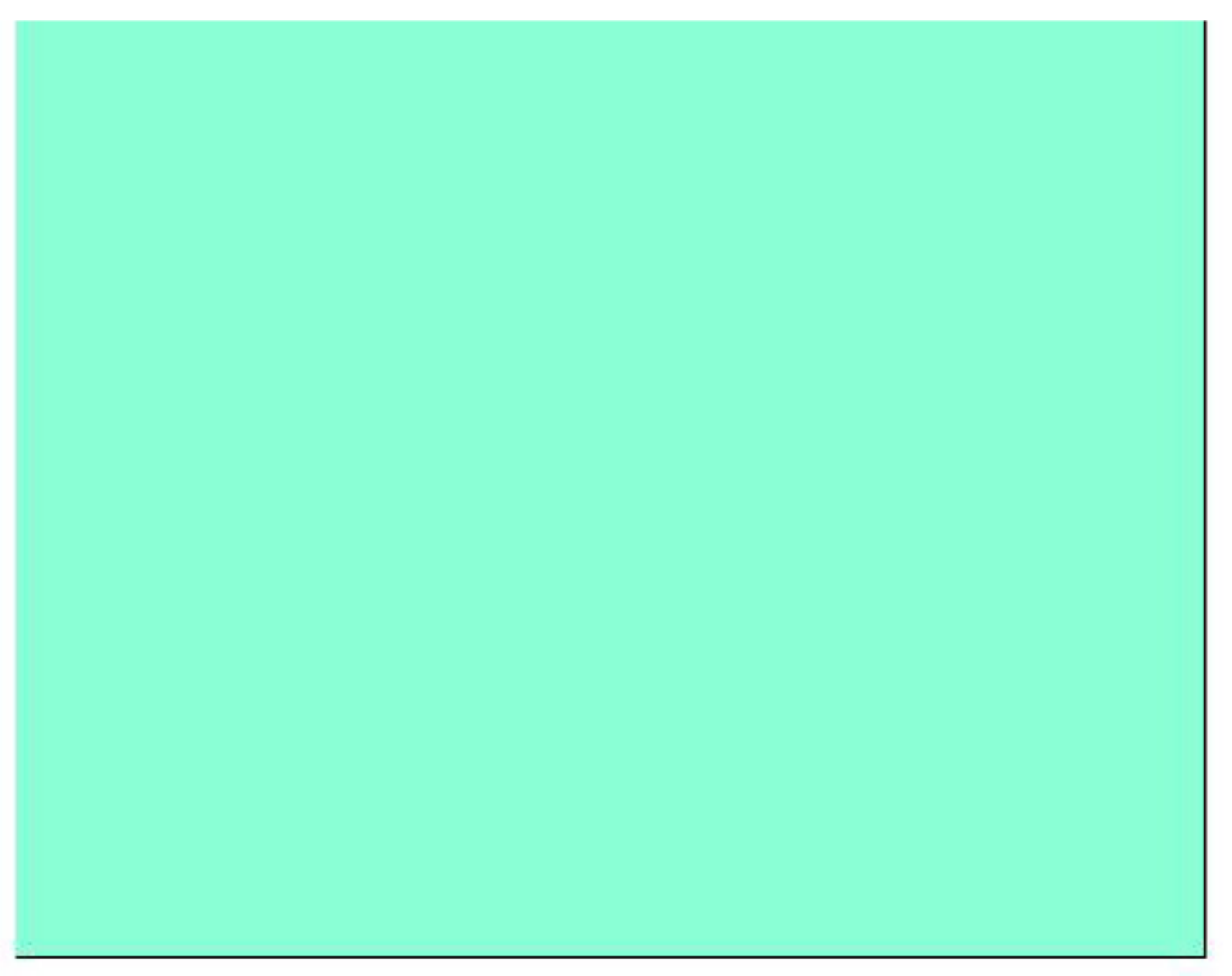}\\ 
		
		180  & (0.487,0.255) & \includegraphics[width=10mm, height=7mm]{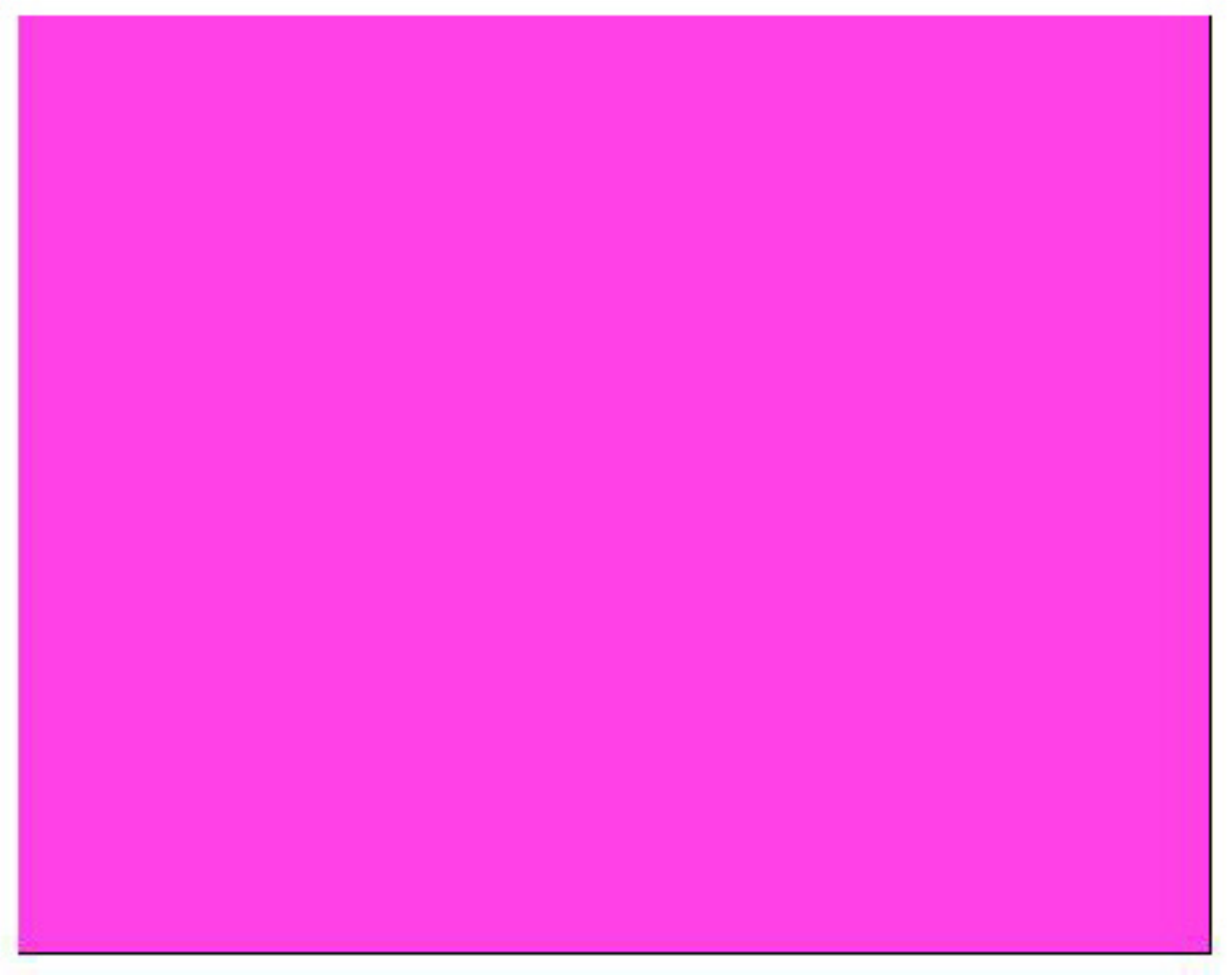} &   380    & (0.299,0.361) & \includegraphics[width=10mm, height=7mm]{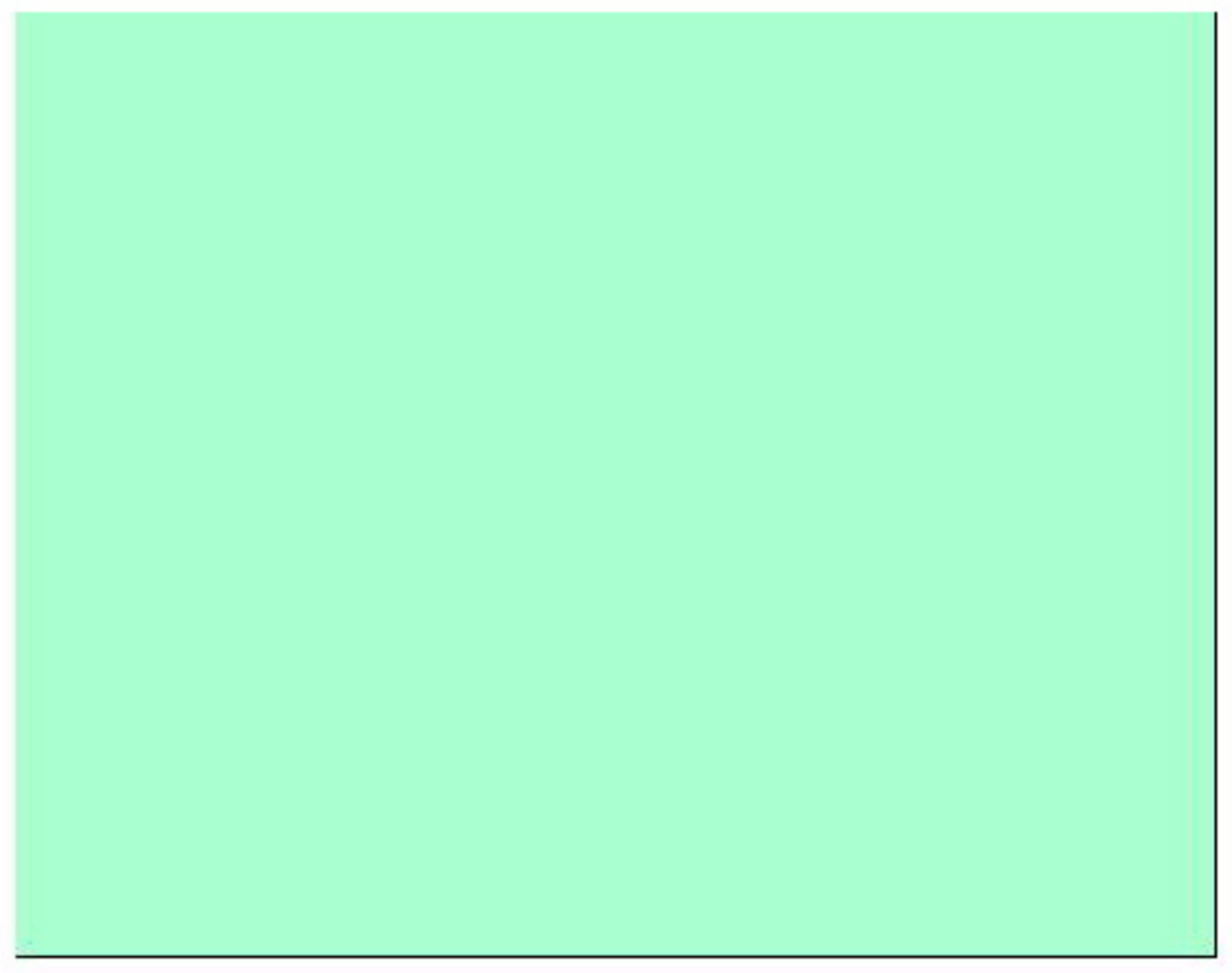}\\    
		
		\hline   
	\end{tabular*}
\end{table*}

Furthermore, the electro-optical response of the proposed structure, when averaged over the graphene membranes, is radius independent. Therefore, changing the radius of the suspended graphene does not result in any qualitative change in the behavior of the pixels \cite{Santiago2018Graphene}. This phenomenon lays the foundation for graphene's mechanical pixels to form high-resolution displays. The RGB responses of graphene pixels are shown in Fig. \ref{Fig4}. The highlighted pixels in the CIE1931 chromaticity space are marked with black dots showing the gamut pixel trajectory in the chromaticity diagram. The simulation results show that with the increase of the number of graphene layers, the obtained color gamut becomes larger, because the graphene optical absorption increases with the increase of the number of graphene layers, thus increasing the modulation amplitude. In order to obtain more abundant colorimetric responses, a thicker membrane is required. The colorimetric response of five-layer graphene is better, with a maximum deflection of about 400 nm. Because this research mainly discusses the influence of ultra-thin graphene on interferometric modulator display technology, only the influence of five-layer of graphene on the structure is discussed. In conclusion, by analyzing the RGB  channels of different layers of graphene mechanical pixels, we obtained the average color range of 2-5 layers of graphene mechanical pixels and their variation with applied voltage. The white dashed line with arrows shows the color trend as the increase of the equilibrium position of graphene deflection, indicating that electro-optical modulation among the three primary colors can be achieved using a single pixel. Due to the consistency of the proposed graphene MEMS with that in Ref. \citenum{Santiago2018Graphene}, the same refresh frequency can also reach up to 400 Hz. In addition, if the superlubricated micromechanical system of graphene is introduced, the modulation speed can be greatly increased and the modulation voltage can be further reduced \cite{2013Observation,ZHOU2020,YIN2020108661}.

The system realizes the rich dynamic modulation of visible light color without introducing complex structure. As shown in Table \ref{tbl}, the equilibrium position of graphene gradually changes from 0 nm to 380 nm (step length interval is 20 nm), and CIE chroma coordinates (x, y) with respect to the visible light transmission spectrum and intuitive color blocks of simulation are given accordingly. As the equilibrium position of graphene gradually increases, the interference color of the transmission spectrum revolve periodically, and the color displayed gradually changes from blue to green and finally to red, covering the whole visible light range. While color saturation gradually decreases due to the influence of parabolic profile of the graphene drums. Since the ideal white color coordinate of the standard illumination D65 is (0.313,0.329) \cite{doi:10.1021/acsnano.9b07523}, the chromaticity coordinates of the interference color of different equilibrium positions of graphene presents the phenomenon of taking points around the light source in the CIE chromaticity diagram. According to the calculation of color block changes, the voltage can be used to manipulate the deflection of graphene, thus verifying the rationality and effectiveness of vivid visible light color realization. It is worth noting that the color rendering system compared with the metarsurface plasma effect can also realize dynamic color adjustment for display devices \cite{doi:10.1021/acsnano.6b08465,doi:10.1063/1.5110051}.

Finally, although it is beyond the scope of this study, the feasibility of the experiment is also discussed. It is noted that the results are obtained through simulation, but all the parameters used in the simulation are based on the experimental results (i.e., the deflection of graphene). Especially, the configuration shown in Fig. \ref{Fig1} can be experimentally achieved following the previous preparation method. On the one hand, the dielectric DBRs were deposited by plasma-enhanced chemical vapor deposition (PECVD) on the epitaxial structure at low temperatures \cite{2002Potential,1250494}. The use of DBRs in combination with a polymer sacrificial layer enables the achievement of the air-cavity  \cite{1250494,2006Fabrication}. On the other hand, multilayer graphene films grown by chemical vapor deposition (CVD) can be transferred into a DBR cavity using a semi-dry transfer technique  \cite{Santiago2018Graphene,doi:10.1021/acs.nanolett.6b02416,doi:10.1021/acs.jpcc.8b10470}. 

\section{Conclusions}
In conclusion, dynamic color modulation in the composite structure of graphene microelectromechanical systems (MEMS)-photonic crystal microcavity is investigated. The designed photonic crystal microcavity has three resonant standing wave modes corresponding to the wavelength of red, green and blue, and the light of these three modes forms the localized light field at different positions of the microcavity. Once graphene is added, it can govern the transmittance of three modes. When graphene is located in the abdomen of the standing wave, which has strong light absorption and therefore the structure's transmittance is low, or shows opposite properties when graphene is located in the node of the standing wave. Thus, we can tune the absorption of light of different colors by voltage regulating the equilibrium position of the graphene MEMS in the microcavity. As a result, we use a single pixel to achieve the output of monochromatic light or multiple mixed colors, greatly expanding the gamut. This method can achieve a high resolution, ultrafast and dynamic color modulation by comparing the color rendering system with the metasurface plasma effect. Our findings reveal the physics behind dynamic color modulation with graphene and may provide guidance for the design and manufacture of ultrahigh resolution, ultrafast modulation and wide color gamut interferometric modulator displays.

\section*{Acknowledgements}
This work was supported by the National Natural Science Foundation of China (NSFC) (Grant No. 11764008, 11964007), the Science and Technology Talent Support Project of the Department of Education in Guizhou Province (Grant No. KY[2018]045), the Science and Technology Foundation of Guizhou Province, China (Grant No. [2020]1Y026), and Construction project of characteristic key laboratory in Guizhou Colleges and Universities (KY[2021]003).

\appendix

\section{Details of the transfer matrix method}

The transfer-matrix method can be used in the calculation to analyze the Bragg mirror micro-cavities \cite{2018Broad,0The,2016Tunable,2019Two}, and the TM (the magnetic field of light paralell to the interface) mode light beam is launched onto the photonic structure. As shown in the Figure \ref{Fig1}, the designed structure is divided into $1,2,3\cdots 2N+5$ layers from top to bottom with the angle of incidence is $\theta=0^{\circ}$, only considered in the case of normal incidence. Derived from Maxwell’s equations and boundary conditions, the incident (reflection) electric field amplitudes $A_{l-1}$($B_{l-1}$) and $A'_{l}$($B'_{l}$) on the up and down sides of the interface can be expressed as\cite{0The,2016Tunable}.
\begin{equation}
\begin{pmatrix}B_{l-1}\\A_{l-1}\end{pmatrix}=\begin{pmatrix}\dfrac{1}{t_l}&\dfrac{r_l}{t_l}\\\dfrac{r_l}{t_l}&\dfrac{1}{t_l}\end{pmatrix}\begin{pmatrix}B'_{l}\\A'_{l}\end{pmatrix}=M_l\begin{pmatrix}B'_{l}\\A'_{l}\end{pmatrix}\label{eqS1}\\
\end{equation}
here, $t_l$ and $r_l$ are the transmission and reflection coefficients, which can be achieved from the Fresnel equations.
$t_l=2n_{l-1}cos\theta_{l-1}/(n_{l-1}cos\theta_l+n_lcos\theta_{l-1})$, and $r_l=(n_{l-1}cos\theta_l-n_lcos\theta_{l-1})/(n_{l-1}cos\theta_l+n_lcos\theta_{l-1})$. $n_l$ is the refractive index of the material in the $l\text{-}th$ layer. If the incident light is the TE (the electric field of light paralell to the interface) mode light, the transmission and reflection coefficients are expressed as $t_l=2n_{l-1}cos\theta_{l-1}/(n_lcos\theta_l+n_{l-1}cos\theta_{l-1})$, and $r_l=(n_lcos\theta_l-n_{l-1}cos\theta_{l-1})/(n_lcos\theta_l+n_{l-1}cos\theta_{l-1})$.

The angle of light propagation in different layers is governed by Snell's Law $n_{l-1}sin\theta_{l-1}=n_lsin\theta_l$. The electric field at two interfaces of the same layer with a thickness of $d_l$ satisfies the transfer equation
\begin{equation}
\begin{pmatrix}B'_l\\A'_l\end{pmatrix}=\begin{pmatrix}e^{-2\pi d_ln_lcos\theta_l/\lambda}&0\\0&e^{2\pi d_ln_lcos\theta_l/\lambda}\end{pmatrix}\begin{pmatrix}B_{l}\\A_{l}\end{pmatrix}=P_l\begin{pmatrix}B_{l}\\A_{l}\end{pmatrix}\label{eqS2}\\
\end{equation}
Where $n_0=n_{2N+5}=1$. The field amplitudes of incoming and outgoing light in the structure can be characterized by the transfer equation
\begin{equation}
\begin{pmatrix}B_0\\A_0\end{pmatrix}=\begin{pmatrix}\prod\limits_{l=1}^{2N+4}M_lP_l\end{pmatrix}M_{2N+5}\begin{pmatrix}B'_{2N+5}\\A'_{2N+5}\end{pmatrix}=\begin{pmatrix}Q_{11}&Q_{12}\\Q_{21}&Q_{22}\end{pmatrix}\begin{pmatrix}B'_{2N+5}\\A'_{2N+5}\end{pmatrix}\label{eqS3}\\
\end{equation}
The transmittance and reflectance of the photonic crystal are $R(\lambda)=\left|Q_{21}/Q_{11}\right|^2$ and $T(\lambda)=\left|1/Q_{11}\right|^2$, respectively.

Although graphene is not exactly parallel to the Bragg mirror, due to the small slope of the graphene working region, we can divide the graphene into many small pieces, each of which can be considered parallel to the Bragg mirror, and the transfer matrix can be used. We can calculate the transmittance of each of the pieces, and then sum them up to get the average transmittance.

\section{Simulation of different graphene placement positions}

\begin{figure}
	\centering
	\includegraphics[width=\linewidth]{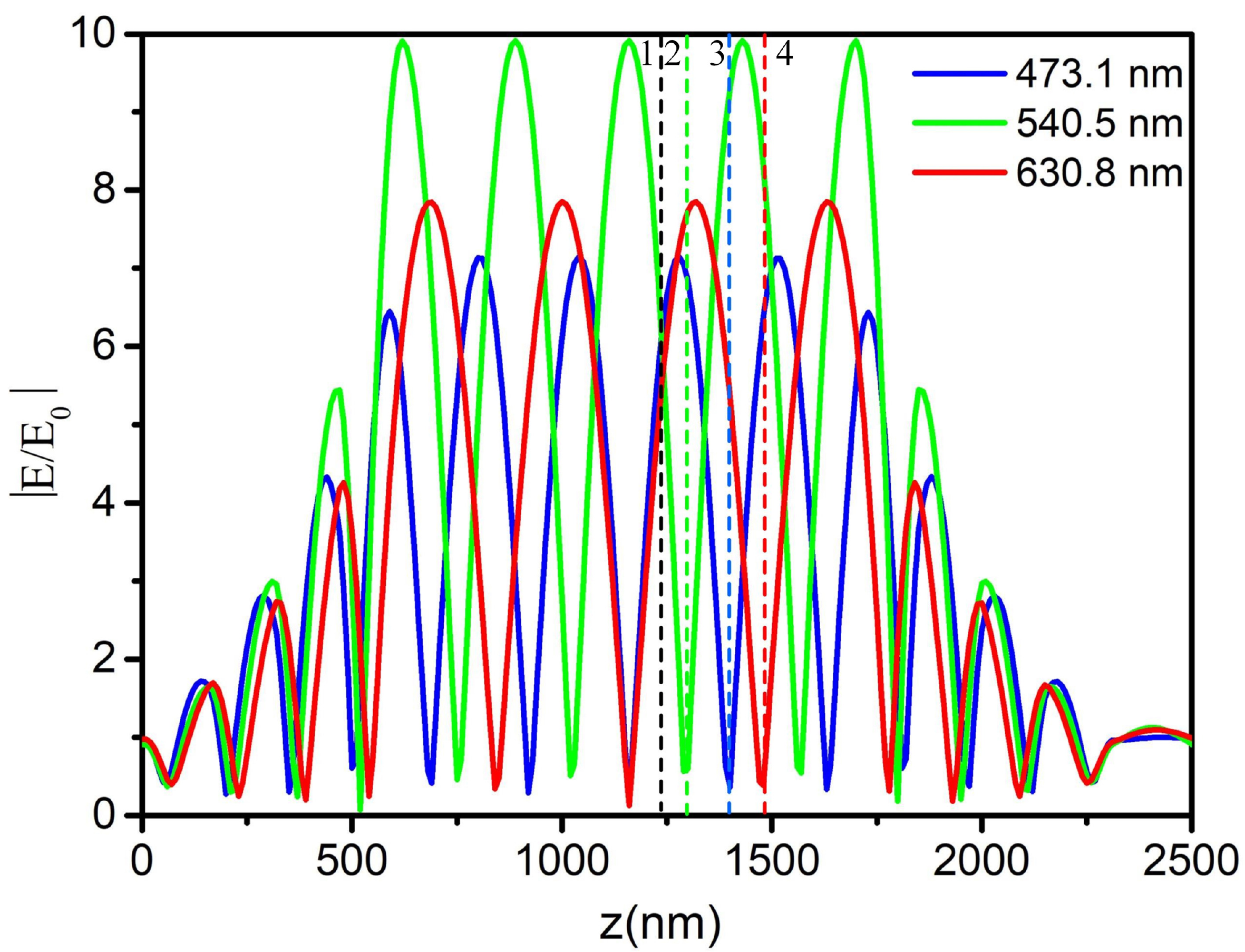}
	\caption{Electric field intensity distribution at the resonant wavelength of the three modes (dashed lines 1-4 are the corresponding positions of 0, 40, 140 and 240 nm in the microcavity in Fig. \ref{Fig6}).}
	\label{Fig5}
\end{figure}

\begin{figure}
	\centering
	\includegraphics[width=\linewidth]{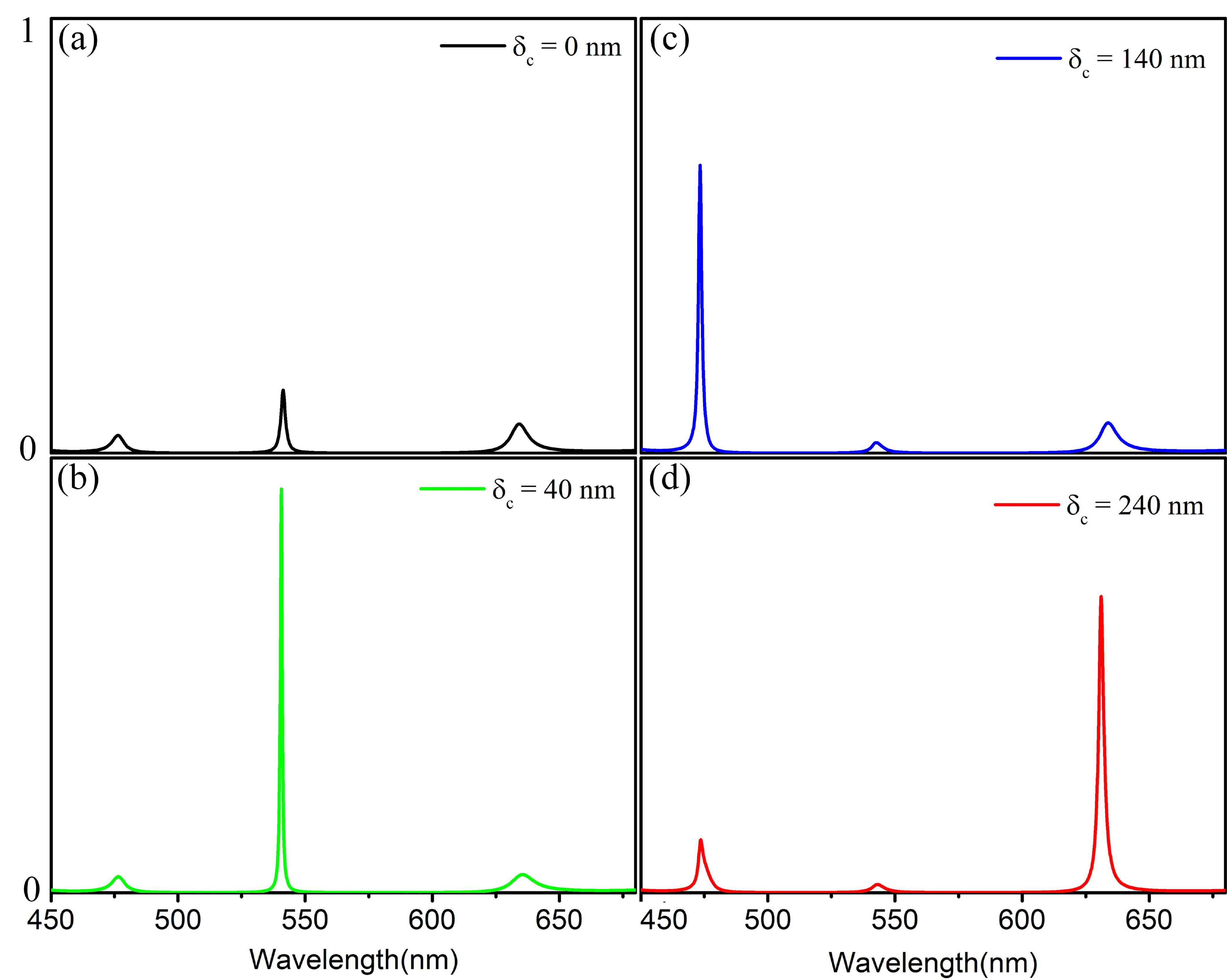}
	\caption{(a)-(d)Transmission spectrum of the system when the equilibrium position of graphene is 0, 40, 140, 240 nm, respectively.}
	\label{Fig6}
\end{figure}

\begin{figure}
	\centering
	\includegraphics[width=\linewidth]{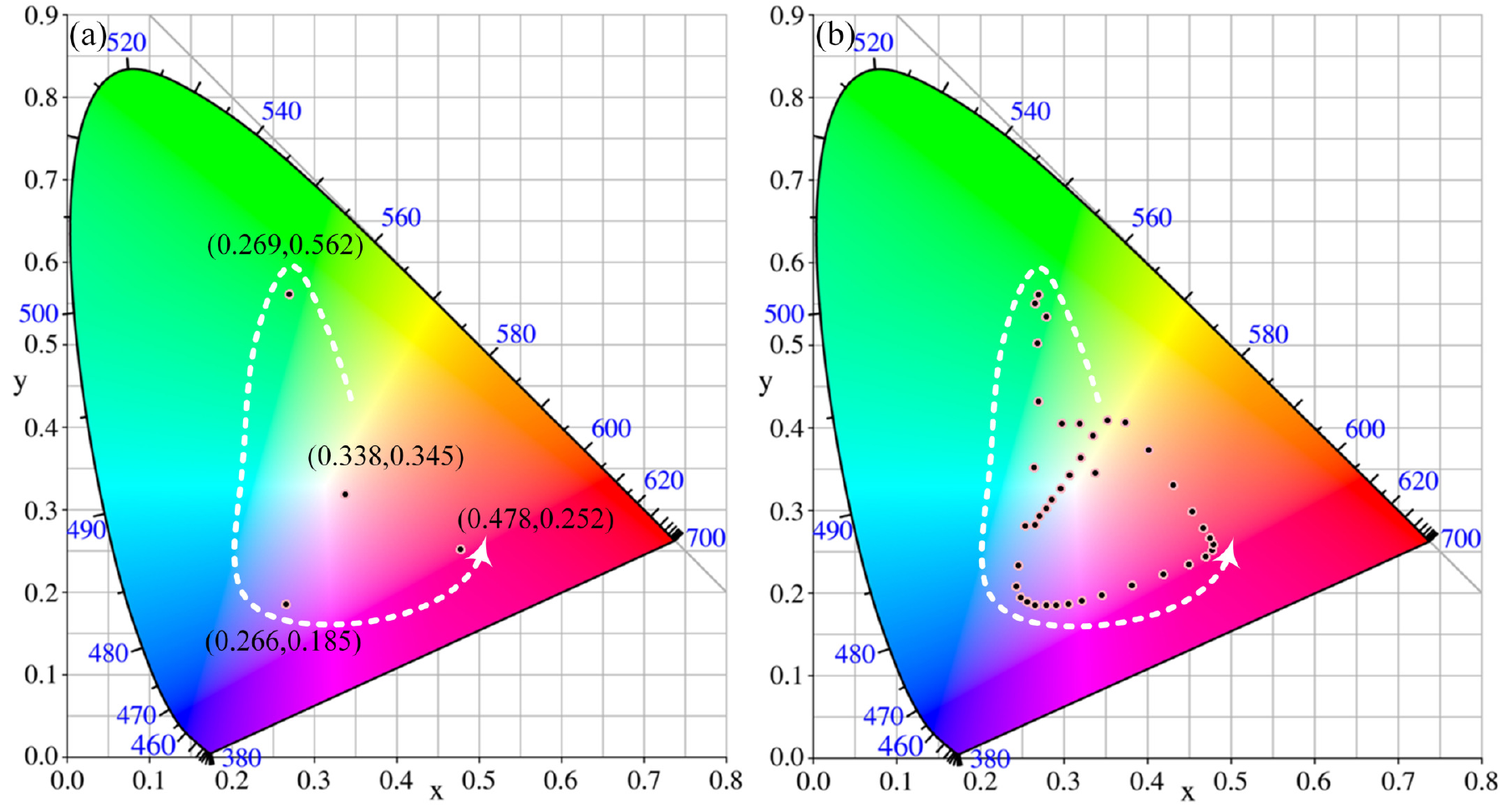}
	\caption{(a) The color coordinates of 0, 40, 140, 240 nm and and the corresponding positions in 1931 CIE chromaticity diagram. (b) Transmission colors with five-layer graphene (the white dashed lines with the arrow represent the evolution trend for the colors as the deflection of graphene increases).}
	\label{Fig7}
\end{figure}

All our simulations are performed using the transfer-matrix method. In the simulations, we consider that the distances between the graphene pixel and the Bragg mirrors are $D_1$ and $D_2$, where $D_1=636$ nm and $D_2=444$ nm, as shown in Fig. \ref{Fig5} the black dashed line. 

A description similar to the text, we take the equilibrium positions of the four cases of graphene extracted are: 0 nm, 40 nm, 140 nm and 240 nm as the research objects. The corresponding dashed line positions 1-4 present in Fig. \ref{Fig5}, and the corresponding transmission spectra are shown in Fig. \ref{Fig6}(a)-(d), respectively.

The color coordinates of the above four cases are located in (0.338, 0.345), (0.269, 0.562) and (0.266, 0.185), and (0.478,0.252) respectively, as shown in Fig. \ref{Fig7}(a). The RGB responses of five-layer graphene pixels are shown in Fig. \ref{Fig7}(b). This fully demonstrates the flexibility of our structural design.

\nocite{*}

\bibliography{apssamp_yanlixu}

\end{document}